\documentclass[onecolumn, journal]{IEEEtran}

\usepackage{cite}

\ifCLASSINFOpdf
  \usepackage[pdftex]{graphicx}
\else
  \usepackage[dvips]{graphicx}
\fi

\usepackage[cmex10]{amsmath}

\usepackage[
    bookmarks=false,        
    pdffitwindow=false,     
    pdfstartview={FitH},    
    colorlinks=true,        
    linkcolor=blue,         
    citecolor=blue,         
    filecolor=blue,         
    urlcolor=blue           
]{hyperref}
\usepackage{amssymb}
\usepackage{nameref}
\usepackage{algorithm}
\usepackage{algorithmicx}
\usepackage{algpseudocode}
\usepackage{multirow}
\usepackage{rotating}

\newcommand{\vect}[1]{\boldsymbol{\mathbf{#1}}}
\DeclareMathOperator*{\argmax}{\mathrm{arg \, max}}
\DeclareMathOperator*{\argmin}{\mathrm{arg \, min}}
\newcommand{\abs}[1]{\lvert#1\rvert}

\begin{document}

\title{A Spatiotemporal Dynamic Solution to the MEG Inverse Problem: An Empirical Bayes Approach}

\author{Camilo~Lamus,
        Matti~S.~H{\"a}m{\"a}l{\"a}inen,
        Simona~Temereanca,
        Emery~N.~Brown,
        and~Patrick~L.~Purdon,
\thanks{C. Lamus is with the Department of Brain and Cognitive Sciences, Massachusetts Institute of Technology, Cambridge, MA 02139 USA (e-mail: camilo@neurostat.mit.edu).}
\thanks{M.S. H{\"a}m{\"a}l{\"a}inen is with Harvard Medical School, Boston, MA 02115 USA, with the Athinoula A. Martinos Center for Biomedical Imaging, Department of Radiology, Massachusetts General Hospital, Charlestown, MA 02129 USA, and with the Department of Neuroscience and Biomedical Engineering, Aalto University School of Science, Espoo, Finland.}
\thanks{S. Temereanca is with Harvard Medical School, Boston, MA 02115 USA, with the Athinoula A. Martinos Center for Biomedical Imaging, Department of Radiology, Massachusetts General Hospital, Charlestown, MA 02129 USA, and with the Department of Brain and Cognitive Sciences, Massachusetts Institute of Technology, Cambridge, MA 02139 USA.}
\thanks{E.N. Brown is with the Department of Brain and Cognitive Sciences and the Institute for Medical Engineering and Science, Massachusetts Institute of Technology, Cambridge, MA 02139 USA, with the Department of Anesthesia, Critical Care and Pain Medicine, Massachusetts General Hospital, Boston, MA 02114 USA, and with Harvard Medical School, Boston, MA 02115 USA.}
\thanks{P.L. Purdon is with Harvard Medical School, Boston, MA 02115 USA, and with the Department of Anesthesia, Critical Care and Pain Medicine, Massachusetts General Hospital, Boston, MA 02114 USA (e-mail: patrickp@nmr.mgh.harvard.edu)}}

\markboth{Preprint,~Vol.~x, No.~y, 2012}%
{Lamus \MakeLowercase{\textit{et al.}}: Dynamic Source Localization}

\maketitle

\begin{abstract}

MEG/EEG are non-invasive imaging techniques that record brain activity with high temporal resolution. However, estimation of brain source currents from surface recordings requires solving an ill-posed inverse problem. Converging lines of evidence in neuroscience, from neuronal network models to resting-state imaging and neurophysiology, suggest that cortical activation is a distributed spatiotemporal dynamic process, supported by both local and long-distance neuroanatomic connections. Because spatiotemporal dynamics of this kind are central to brain physiology, inverse solutions could be improved by incorporating models of these dynamics. In this article, we present a model for cortical activity based on nearest-neighbor autoregression that incorporates local spatiotemporal interactions between distributed sources in a manner consistent with neurophysiology and neuroanatomy. We develop a dynamic Maximum a Posteriori Expectation-Maximization (dMAP-EM) source localization algorithm for estimation of cortical sources and model parameters based on the Kalman Filter, the Fixed Interval Smoother, and the EM algorithms. We apply the dMAP-EM algorithm to simulated experiments as well as to human experimental data. Furthermore, we derive expressions to relate our dynamic estimation formulas to those of standard static models, and show how dynamic methods optimally assimilate past and future data. Our results establish the feasibility of spatiotemporal dynamic estimation in large-scale distributed source spaces with several thousand source locations and hundreds of sensors, with resulting inverse solutions that provide substantial performance improvements over static methods.

\end{abstract}


\IEEEpeerreviewmaketitle

\section{Introduction}
\label{sec:intro}

Magnetoencephalography (MEG) and electroencephalography (EEG) are non-invasive brain imaging techniques that provide high temporal resolution measurements of magnetic and electric fields at the scalp generated by the synchronous activation of neuronal populations. It has been estimated that a detectable signal can be recorded if as few as one in a thousand synapses become simultaneously active in an area of about 40 square millimeters of cortex~\cite{Hamalainen:1993ws}. The exceptional time resolution of these techniques provides a unique window into the dynamics of neuronal process that cannot be obtained with other functional neuroimaging modalities, such as functional magnetic resonance imaging (fMRI) and positron emission tomography (PET), which measure brain activity indirectly through associated slow metabolic or cerebrovascular changes.

Localizing active regions in the brain from MEG and EEG data requires solving the neuromagnetic inverse problem, which consists of estimating the cerebral current distribution underlying a time series of measurements at the scalp. The ill-posed nature of the electromagnetic inverse problem and the relatively large distance between the sensors and the sources limit the spatial resolution of MEG and EEG. For the inverse problem, solution of the corresponding forward problem in MEG/EEG, i.e., determining the measured magnetic and electric field at the scalp generated by a given distribution of neuronal currents, is a prerequisite. This problem can be solved by adopting a quasistatic approximation to Maxwell's equations, resulting in a linear map which relates the activity of arbitrarily located neuronal sources to the signals in a set of sensors~\cite{Hamalainen:1993ws,Mosher:1999vf}.

Two types of models have been proposed in the neuromagnetic inverse problem literature: equivalent current dipole models and distributed source models. Equivalent current dipole models are based on the assumption that a small set of current dipoles with unknown locations, amplitudes, and orientations can closely approximate the actual current distribution~\cite{Mosher:1992ij,Uutela:1998wr,Bertrand:2001vt,Somersalo:2003wb,Jun:2005bz,Kiebel:2008fa,Campi:2008br}. Distributed source models, on the other hand, assume that the recorded activity results from the activation of a spatial distribution of current dipoles with known locations (see~\cite{Baillet:2001uo} for a review). Most source localization methods assume that scalp measurements and their underlying sources are independent across time, and convenient probabilistic or computational prior constraints are imposed to obtain unique solutions. For example, inverse methods have been proposed that penalize current sources with large amplitude or power~\cite{Hamalainen:1994uk,VanVeen:1997tn,Uutela:1999ca,Auranen:2005cm}, impose spatial smoothness~\cite{PascualMarqui:1994tp}, or favor focal estimates~\cite{Gorodnitsky:1995wv}. Furthermore, Bayesian methods have been used to obtain spatially sparse estimates, where components of the current source covariance are estimated directly~\cite{Phillips:2005hd,Mattout:2006cq} or additional hierarchical priors are assigned in order to compute posterior distributions~\cite{Sato:2004gp,Nummenmaa:2007gq,Wipf:2009gj}. The assumption of temporal independence in all of these methods allows the inverse solution at each point in time to be computed individually, without regard for dynamics, treating the probability distribution of the underlying sources as static in time. While this approach is computationally convenient, it ignores the temporal structure observed in neural recordings at many different levels~\cite{Buzsaki:2004ut}, which could be used to improve inverse solutions.

Recent methods for source localization have incorporated temporal smoothness constraints as part of a general Bayesian framework. The approach taken in these methods is to specify an arbitrary prior distribution for the dipole sources in space and time, sometimes in terms of basis functions, with limited or space-time separable interactions in order to obtain simplified estimation algorithms~\cite{Baillet:1997ue,Greensite:2003wn,Daunizeau:2006wl,Daunizeau:2007ji,Friston:2008jr,TrujilloBarreto:2008ck,Zumer:2008in,Limpiti:2009ca,Ou:2009cm,Bolstad:2009eg}. Linear state-space models have also been used to model source current dynamics. However, in order to reduce computational complexity, these methods either apply spatially-independent approximations to their respective estimation algorithms~\cite{Yamashita:2004fb,Galka:2004ef} or \textit{a priori} fix model specific parameters to avoid the problem of parameter estimation~\cite{Long:2011fn}. Overall, while these methods incorporate temporal structure in their models, they specify highly constrained spatiotemporal interactions, such as space-time separability or spatial independence, that may not accurately reflect dynamic relationships between different brain areas.

Converging lines of evidence from neurophysiology, biophysics, and neuroimaging illustrate that dynamic spatiotemporal interactions are a central feature of brain activity. Intracranial recordings in different species, including humans, exhibit strong spatial correlations during rest and task periods that persist up to distances of 10 mm along the cortical surface~\cite{Bullock:1995wy,Destexhe:1999tc,Leopold:2003va}. These local spatial interactions are supported neuroanatomically by axonal collateral projections from pyramidal cells that spread laterally at distances of up to 10 mm~\cite{Nunez:1995vc}. Biophysical spatiotemporal dynamic models of neuronal networks at various levels of abstraction have been effective in reproducing properties of electromagnetic scalp recordings seen during both normal and disease states~\cite{Jirsa:2002wz,Wright:2004wl,Robinson:2005dw,David:2005fx,Sotero:2007ij,Kim:2007be,Izhikevich:2008ej,Gross:2001cc}. Furthermore, fMRI and PET studies have shown temporally coherent fluctuation in activation within widely distributed cortical networks during resting-state and experimentally administered task periods~\cite{Raichle:2001vf,Gusnard:2001ut,Fox:2005kx,Fox:2007ig}.

In this article we present a new dynamic source localization method that models local spatiotemporal interactions between distributed cortical sources in a manner consistent with neurophysiology and neuroanatomy, and then uses this model to estimate an inverse solution. Specifically, 1) we describe a model of the spatiotemporal dynamics based on nearest-neighbors multivariate autoregression along the cortical surface; 2) We develop an algorithm for dynamic estimation of cortical current sources and model parameters from MEG/EEG data based on the Kalman Filter, the Fixed Interval Smoother, and the Expectation-Maximization (EM) algorithms; 3) We derive expressions to relate our dynamic estimation formulas to those of standard static algorithms; and 4) We apply our spatiotemporal dynamic method to simulated experiments of focal and distributed cortical activation as well as experimental data from a human subject.

\section{Methods}
\label{sec:methods}

\subsection{Measurement model}
\label{subsec:measmodel}

In an MEG/EEG experiment, we obtain a temporal set of recordings from hundreds of sensors located above the scalp. The data are sampled by $n$ sensors at times $\{\Delta t\}_{t=1}^T$, where $\Delta$ is the sampling interval and $T$ is the number of measurements in time. Let $y_{i,t}$ denote the measurement at time $t$ in sensor $i$, and define ${\vect{y}_t = [y_{1,t}, y_{2,t}, \dots, y_{n,t}]'}$ as the $n \times 1$ vector of measurements at all sensors at time $t$. We assume that the measurements were generated by $p$ current dipole sources distributed on the cortical surface and oriented perpendicular to it. Let $\beta_{i,t}$ denote the source amplitude of the $i$th dipole at time $t$, and define ${\vect{\beta}_t = [\beta_{1,t}, \beta_{2,t}, \dots, \beta_{p,t}]'}$ as the $p \times 1$ vector of cortical source activity in all considered locations, or cortical state vector, at time $t$. Typically, $n$ $\sim$ a few hundred, and $p$ $\sim$ several thousand. The relationship between the measurement vector $\vect{y}_t$ and the cortical state vector $\vect{\beta}_t$ is given by the measurement equation,

\begin{equation} \label{eq:measmodel}
    \vect{y}_t = \vect{X}\vect{\beta}_t + \vect{\varepsilon}_t,
\end{equation}

\noindent where $\vect{X}$ is the $n \times p$ lead field gain matrix computed using a quasistatic approximation to Maxwell's equations~\cite{Hamalainen:1993ws}, i.e., the solution of the forward problem, and $\vect{\varepsilon}_t$ is a $n \times 1$ Gaussian white noise vector with zero mean covariance matrix equal to the identity matrix $\vect{I}$ and independent from $\vect{\beta}_t$ for all time points. In Equation~\eqref{eq:measmodel} we assumed that the model has been spatially whitened, i.e., that the original raw data model ${\tilde{\vect{y}}_t = \tilde{\vect{X}}\vect{\beta}_t + \tilde{\vect{\varepsilon}}_t}$ has been premultiplied by the inverse of a matrix square root of the covariance of $\tilde{\vect{\varepsilon}}_t$. Since the orientation of the current generators of the electromagnetic field, i.e., the apical dendrites of pyramidal cells, is perpendicular to the cortical surface, the choice of fixing the current dipole orientation along this direction is justified~~\cite{Hamalainen:1993ws}. Nevertheless, our development can be easily extended to account for unconstrained source orientations.

\subsection{Spatiotemporal dynamical source model}
\label{subsec:spacetimesourcemodel}

As we pointed in the Introduction Section \ref{sec:intro}, evidence from neurophysiology studies, neuroanatomy, biophysics, and neuroimaging suggests that cortical activation is a distributed spatiotemporal dynamic process~\cite{Bullock:1995wy,Destexhe:1999tc,Leopold:2003va,Nunez:1995vc,Jirsa:2002wz,Wright:2004wl,Robinson:2005dw,David:2005fx,Sotero:2007ij,Kim:2007be,Izhikevich:2008ej,Gross:2001cc,Raichle:2001vf,Gusnard:2001ut,Fox:2005kx,Fox:2007ig}. Because spatiotemporal dynamics of this kind are fundamental to brain physiology, inverse solutions could be greatly improved by incorporating models that approximate these dynamics.

One way to model local spatiotemporal connections of this type is to use a nearest-neighbor autoregressive model. In this autoregressive model, neuronal currents at a given point in time and space $\beta_{i,t}$ are a function of past neuronal currents within a small local neighborhood along the cortical surface and a small perturbation $\omega_{i,t}$ that accounts for unknown factors affecting the evolution of cortical currents:

\begin{equation} \label{eq:nearestneigh}
    \beta_{i,t} = \phi [ \underbrace{f_{i,i} \beta_{i,t-1}}_{\text{Past activity}} + \underbrace{\sum_{j \in \mathcal{N}(i)} f_{i,j} \beta_{j,t-1}}_{\text{Past activity of neighbors}}] + \underbrace{(1-\phi^2)^{1/2}\omega_{i,t}}_{\text{Unaccounted factors}}.
\end{equation}

\noindent In Equation~\eqref{eq:nearestneigh} $\mathcal{N}(i)$ represents the index set of nearest neighbors of the $i$th dipole source, and $\phi$ (${0\leq \phi < 1}$) is a scalar that both represent the strength of the history dependence in the dynamics and also guarantee the stability of the cortical state dynamics. The state noise ${\vect{\omega}_t = [\omega_{1,t}, \omega_{2,t}, \dots, \omega_{p,t}]}$ is a $p \times 1$ Gaussian vector with zero mean and covariance matrix $\vect{Q}$, independent from the measurement noise $\vect{\varepsilon}_t$. The weights $f_{i,j}$, which represent the individual influence of neighboring sources in the autoregression, are assumed to decay with the distance from the $i$th to the $j$th dipole source. A simple relation that reflects this modeling assumption is to make the weights inversely proportional to the distance between dipoles,

\begin{equation} \label{eq:invdistance}
    f_{i,j} \propto \frac{1}{\text{distance from $i$th to $j$th dipole}},
\end{equation}

\noindent where the proportionality constant is chosen such that the contribution of the neighbors to the dynamics of the ${i}$th source equals its self contribution, while the total contribution is equal to one: ${\sum_{j\in \mathcal{N}(i)}f_{i,j} = f_{i,i}}$ and ${\sum_{j\in \mathcal{N}(i)}f_{i,j} + f_{i,i} = 1}$. This allows the power (prior variance) in every dipole $\beta_{i,t}$ to remain approximately constant over time.

\begin{figure}[htb]
    \begin{center}
    \includegraphics{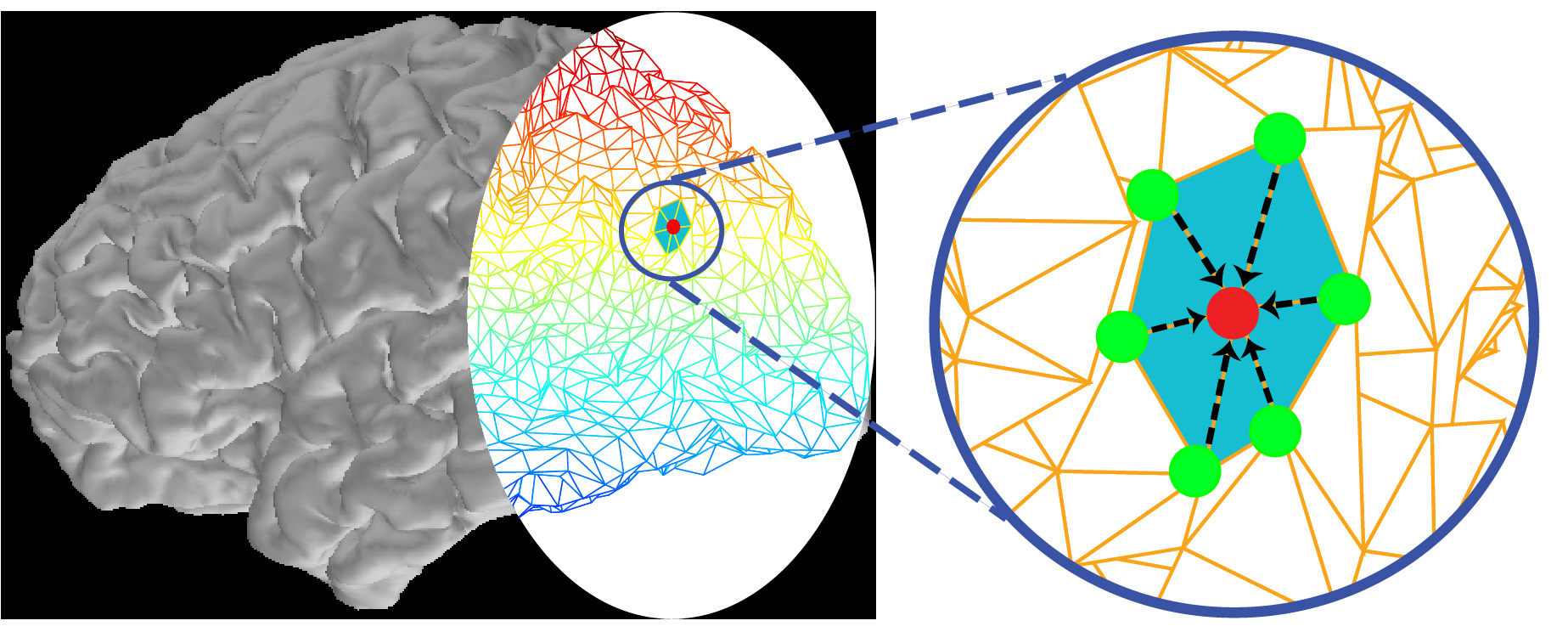}
    \end{center}
    \caption{{\bf Illustration of the spatiotemporal dynamic source model}. The left panel shows the reconstructed cortical surface where the caudal portion of cortex is depicted as a triangulated mesh of dipole sources. The zoomed-in panel in the right shows a dipole source (central red dot) and its nearest-neighbor dipole sources (green dots). At a given point in time, the activity at the central dipole (red dot) is a function of its past activity and the weighted activity of a small neighborhood of dipole sources (green dots), plus a perturbation that represents unknown factors affecting the dipole source activity. The weights (black dashed arrows) are inversely proportional to the distance from the central dipole.}
    \label{fig:illusourcemodel}
\end{figure}

Figure \ref{fig:illusourcemodel} illustrates this nearest-neighbors autoregressive model. The left panel shows the cortical surface reconstructed from high-resolution magnetic resonance images (MRI) using \textit{Freesurfer}~\cite{Dale:1999ks,Fischl:1999va}, where the caudal portion is represented by its triangulated mesh of dipoles. The zoomed-in panel in the right isolates a dipole source (central red dot) and its corresponding nearest-neighbor dipoles (green dots). At a given point in time, the activity of the central dipole (red dot) is a function of its past activity and the weighted activity of a small neighborhood of dipole sources (green dots), where the weights (black dashed arrows) are inversely proportional to the distance from the central dipole to its neighbor. We employ a cortical surface representation with $p=5124$ dipoles sources, with an average distance of $6.2$ mm between nearest neighbors, yielding a model that is consistent with the local spatial properties of intracranial electrophysiology and neuroanatomy. In Appendix \ref{app:robustnessmodelF}, we analyze how modifications in our spatial model, such us those arising from choosing a different weighting scheme or a misspecification of the weights $f_{i,j}$ (Eq. \ref{eq:nearestneigh}), influence the smoothness encoded \textit{a priori} in the dipole covariance. There we show that as long as the modification in the spatial model is not too large, the smoothness modeled in the dipole sources is not dramatically altered.

We can readily rewrite Equation \eqref{eq:nearestneigh} as the multivariate autoregressive model,

\begin{equation} \label{eq:statespace}
    \vect{\beta}_t = \phi \vect{F}\vect{\beta}_{t-1} + (1-\phi^2)^{1/2}\vect{\omega}_t,
\end{equation}

\noindent where the state feedback matrix $\vect{F}$ encompasses the neighborhood interactions between the sources at time $t$ in terms of the sources in the previous time step $t-1$:

\begin{equation} \label{eq:statefeedbackmat}
    (\vect{F})_{i,j} = \begin{cases}
                    f_{i,j}    & \text{if $j=i$ or $j \in \mathcal{N}(i)$} \\
                    0          & \text{otherwise}
                \end{cases},
\end{equation}

\noindent and $(\vect{F})_{i,j}$ is the $i$th row of the $j$th column of $\vect{F}$. The initial dipole source vector $\vect{\beta}_0$ is assumed to be Gaussian with zero mean and covariance matrix $\vect{C}_0$, and independent of $\{\vect{\omega}_t\}_{t=1}^T$.

\subsection{Prior model for the parameters}
\label{subsec:priormodel}

The model defined in Sections \ref{subsec:measmodel} and \ref{subsec:spacetimesourcemodel} yields a probabilistic ``forward model'' for the dynamic generation of the electromagnetic scalp recording. The next step in the description of our model is to define prior densities for the unknown parameters $\{\vect{X},\vect{F},\vect{Q},\vect{C}_0\}$. We should note that while in theory all the parameters in our model have uncertainties, some of them can be fixed \textit{a priori} based on knowledge of the system under investigation. For example, the lead field matrix $\vect{X}$ can be computed using the boundary-element model based on high-resolution magnetic resonance images (MRI)~\cite{Hamalainen:1989ec}. The covariance $\vect{C}_0$ can be updated as described in Section \ref{subsec:dmapemalgorithm}. The state feedback matrix $\vect{F}$ is constructed as indicated in Section \ref{subsec:spacetimesourcemodel} to incorporate nearest-neighbor interaction in order to model basic local intracortical dynamic connections. The selection of the values of $\phi$ (Eq. \ref{eq:statespace}) is discussed in Section \ref{subsec:inialgor}.

We note that the process of fixing the lead field $\vect{X}$ and state feedback $\vect{F}$ matrices can be seen as the analog of selecting the matrix of covariates or regressors in a linear model, where the experimenter's knowledge guides the selection of the regressor as means of testing a scientific hypothesis based on collected data. In our case, the regressors in $\vect{F}$, which are fixed based on neurophysiological modeling considerations, determine how much the state $\vect{\beta}_t$ can be explained by its past $\vect{\beta}_{t-1}$, and the unexplained portion is left to the state noise term $\vect{\omega}_t$. In the same way, the regressors in $\vect{X}$, obtained from geometrical and biophysical knowledge, separate the measured MEG $\vect{y}_t$ between the signals coming from the brain $\vect{\beta}_t$, and the instrument and environmental noise $\vect{\varepsilon}_t$.

The remaining unknown parameters, which must be estimated from data, are the elements of the state noise covariance matrix $\vect{Q}$. We assume this matrix is diagonal, and parametrize it as follows to more clearly interpret parameter estimates:

\begin{equation} \label{eq:statenoisecov}
    \vect{Q}(\vect{\nu}) = [\lambda\mathrm{tr}(\vect{\widehat{\Sigma}})]^{-1}\mathrm{diag}(\vect{\nu}).
\end{equation}

\noindent In Equation~\eqref{eq:statenoisecov} ${\vect{\widehat{\Sigma}} = \vect{X}'\vect{X}/n}$ is the sample covariance of the rows of $\vect{X}$, $\lambda>0$ is equal to the inverse of the signal-to-noise ratio ($\mathrm{SNR}^2$) of the data\footnote{Specifically, if the matrix $\vect{F}$ is set equal to the identity matrix, and ${\nu_i = 1}$ (${i=1, \ldots, p}$), the steady state covariance of $\vect{\beta_t}$ becomes ${\vect{C} = \lim_{t \rightarrow \infty} \mathrm{Cov}(\vect{\beta}_t) = [\lambda\mathrm{tr}(\vect{\widehat{\Sigma}})] ^ {-1}\vect{I}}$\cite{Kailath:2000vg}. If we define the power signal-to-noise ratio as ${\mathrm{SNR}^2 = \mathrm{E}||\vect{X}\vect{\beta}_t||^2 / \mathrm{E}||\vect{\varepsilon}_t||^2}$, then ${\mathrm{SNR}^2 = \mathrm{tr}(\vect{X}'\vect{X})[\lambda\mathrm{tr}(\vect{\widehat{\Sigma}})] ^ {-1} / n = 1/\lambda}.$}, and ${\vect{\nu} = [\nu_1, \nu_2, \dots, \nu_p]'}$ ${(\nu_i>0)}$ is the parameter vector we need to estimate. We should note that the number of variance parameters we need to estimate is in the order of thousands ($p \approx 5000$).

A common probability model for the parameter vector $\vect{\nu}$ is the conjugate prior distribution. In this case the conjugate prior follows an inverse gamma density:

\begin{equation} \label{eq:priordensity}
    \mathrm{Pr}(\vect{\nu}) = \prod_{i=1}^{p} \frac{b^{b-1}}{\Gamma(b-1)} \left(\frac{1}{\nu_i}\right)^{b} \exp\left(\frac{-b}{\nu_i}\right)
\end{equation}

\noindent where the hyperparameter $b$ can be set to a value slightly higher than $3$ to make the mode of this prior close to $1$, and thus consistent with the order of magnitude in the model units, and at the same time maximize its variance yielding a non-informative prior (see Appendix \ref{app:noninfprior}).

\subsection{Empirical Bayes inference with the dynamic Maximum a Posteriori Expectation-Maximization algorithm (dMAP-EM)}
\label{subsec:dmapemalgorithm}

In order to localize active regions of cortex from scalp recordings, we have to derive estimates for the sequence of source amplitudes

\begin{equation} \label{eq:seqdipolesource}
    \{\vect{\beta}_t\}_{t=1}^T = \{\vect{\beta}_1, \vect{\beta}_2, \ldots, \vect{\beta}_t\}
\end{equation}

\noindent and parameters

\begin{equation} \label{eq:params}
    \vect{\nu} = [\nu_1, \nu_2, \ldots, \nu_{p}]'
\end{equation}

\noindent in the model. Specifically, we have to find the Maximum a Posteriori (MAP) estimate of the parameters,

\begin{equation} \label{eq:mapestimateparam}
    \vect{\hat{\nu}}_{\textsc{map}} = \argmax_{\vect{\nu}} \mathrm{Pr}(\vect{\nu} | \{\vect{y}_t\}_{t=1}^T),
\end{equation}

\noindent where $\mathrm{Pr}(\vect{\nu} | \{\vect{y}_t\}_{t=1}^T)$ is the posterior density of the parameters $\vect{\nu}$ conditioned on the full set of measurements

\begin{equation} \label{eq:seqmeas}
    \{\vect{y}_t\}_{t=1}^T = \{\vect{y}_1, \vect{y}_2, \ldots, \vect{y}_T\},
\end{equation}

\noindent and the Empirical Bayes estimate of the source amplitudes, i.e., the conditional mean of the state vector given the full set of measurements $\{\vect{y}_t\}_{t=1}^T$ and the estimate of the parameters in the model $\vect{\hat{\nu}}_{\textsc{map}}$,

\begin{equation} \label{eq:empiricalbayesest}
    \vect{\beta}_{t | T} = \mathrm{E}(\vect{\beta}_t | \{\vect{y}_t\}_{t=1}^T, \vect{\hat{\nu}}_{\textsc{map}}).
\end{equation}

\noindent In Equation~\eqref{eq:empiricalbayesest}, the notation in the subscript of $\vect{\beta}_{t | T}$ indicates that we are conditioning on the measurements form time $1$ until time $T$.

Once we obtain the MAP estimate of the parameters $\vect{\hat{\nu}}_{\textsc{map}}$, computing the expectation in the Empirical Bayes estimate of the source amplitudes (Eq. \ref{eq:empiricalbayesest}) can be readily obtained using the well-known Kalman Filter~\cite{Kalman:1960kh} and Fixed Interval Smoother algorithms~\cite{Rauch:1965wx}. However, the direct optimization required to compute $\vect{\hat{\nu}}_{\textsc{map}}$ (Eq. \ref{eq:mapestimateparam}) would be computationally intractable. Therefore, we developed a dynamic Maximum a Posteriori Expectation-Maximization (dMAP-EM) algorithm to estimate source amplitudes and parameters in our model (Eqs. \ref{eq:mapestimateparam} and \ref{eq:empiricalbayesest}) based on the work of~\cite{Shumway:1982wt} and~\cite{Green:1990ta}. The Expectation-Maximization algorithm~\cite{Dempster:1977ul} is a general iterative method to obtain Maximum Likelihood or Maximum a Posteriori estimates when the observed data can be regarded as incomplete. In our case we treat the sequence of measurement and dipole sources, $\{\vect{y}_t\}_{t=1}^T,\,\{\vect{\beta}_t\}_{t=0}^T$ as the \textit{complete data}, and iterate performing an E-step followed by an M-step until convergence is achieved. In each iteration of the algorithm, the expectations in the E-step (Section \ref{subsubsec:estep}) can be computed with the Kalman Filter~\cite{Kalman:1960kh}, Fixed Interval Smoother~\cite{Rauch:1965wx}, and lag-covariance~\cite{deJong:1988vu} recursions, and the maximization in the M-step (Section \ref{subsubsec:mstep}) can be obtained in closed form. In each iteration we can evaluate the posterior density of the parameters $\vect{\nu}$ using the \textit{innovations} form~\cite{Kitagawa:1996p85} (Section \ref{subsubsec:evalogpost}).

\subsubsection{E-step}
\label{subsubsec:estep}

The dMAP-EM algorithm is initialized with the parameters $\vect{Q}(\vect{\nu}^{(0)})$ and $\vect{C}_0^{(0)}$. In the $i$th iterate of the E-step, the algorithm computes the conditional expectation of the complete data log-likelihood $\mathrm{Pr}(\{\vect{y}_t\}_{t=1}^T,\{\vect{\beta}_t\}_{t=0}^T | \vect{\nu})$, given the observed data $\{\vect{y}_t\}_{t=1}^T$ and the estimate of the parameters $\vect{\nu}^{(i)}$, with an added term for the log-prior density:

\begin{equation} \label{eq:estep}
    U(\vect{\nu} | \vect{\nu}^{(i)})  = \mathrm{E} \left[\log \mathrm{Pr}\left(\{\vect{y}_t\}_{t=1}^T,\{\vect{\beta}_t\}_{t=0}^T | \vect{\nu} \right)
                                    | \{\vect{y}_t\}_{t=1}^T, \vect{\nu}^{(i)}\right] + \log \mathrm{Pr}(\vect{\nu}).
\end{equation}

\noindent We emphasize that the expectation in Equation \eqref{eq:estep} is computed with respect to $\mathrm{Pr}(\{\vect{\beta}_t\}_{t=0}^T | \{\vect{y}_t\}_{t=1}^T, \vect{\nu}^{(i)})$, where we have conditioned on the full set of measurements and the parameter estimate of the previous iteration.

Before continuing with the algorithm we establish the following variables to simplify the notation. We define the conditional mean

\begin{equation} \label{eq:condmean}
    \vect{\beta}_{t | \tau}^{(i)} = \mathrm{E}(\vect{\beta}_t | \{\vect{y}_j\}_{j=1}^{\tau}, \vect{\nu}^{(i)}),
\end{equation}

\noindent the conditional covariance

\begin{equation} \label{eq:condcov}
    \vect{V}_{t | \tau}^{(i)} = \mathrm{Cov}(\vect{\beta}_t | \{\vect{y}_k\}_{j=1}^{\tau}, \vect{\nu}^{(i)}),
\end{equation}

\noindent and the conditional lag-covariance

\begin{equation} \label{eq:condlagcov}
    \vect{V}_{t,t-1 | \tau}^{(i)} = \mathrm{Cov}(\vect{\beta}_t,\vect{\beta}_{t-1} | \{\vect{y}_j\}_{j=1}^{\tau}, \vect{\nu}^{(i)}),
\end{equation}

\noindent where the subscript in Equations \eqref{eq:condmean}, \eqref{eq:condcov}, and \eqref{eq:condlagcov} indicate that we are conditioning on the measurements form time $1$ until time $\tau$, and the superscript $i$ refers to the iteration number. For example, as we used in Equation \eqref{eq:empiricalbayesest}, by setting $\tau$ equal the number of samples in time $T$ we indicate that are conditioning on the full set of measurements.

As shown in Appendix \ref{app:derestep}, the computation of the conditional expectations in the function $U(\vect{\nu} | \vect{\nu}^{(i)})$ (Eq. \ref{eq:estep}) can be obtained in closed form yielding

\begin{align} \label{eq:ufunction}
U(\vect{\nu} | \vect{\nu}^{(i)}) &=
    -\frac{1}{2} \left\{ c_{1}+\log |\vect{C}_0|
        + \mathrm{tr}\left[\vect{C}_0^{-1}
        \left(\vect{V}_{0 | T}^{(i)}+ \vect{\beta}_{0 | T}^{(i)}\vect{\beta}^{(i)}_{0 | T}\right)\right] \right\}   \nonumber \\
    &\quad - \frac{1}{2} \left\{ c_{2}T + T\log|\vect{Q}(\vect{\nu})|
        + (1-\phi^2)^{-1} \mathrm{tr} \left[ \vect{Q}(\vect{\nu})^{-1}\vect{A}^{(i)} \right] \right\}   \nonumber \\
    &\quad - \frac{1}{2} \left\{ c_{3}T
        + \mathrm{tr} \left[\vect{B}^{(i)} \right] \right\} \nonumber \\
    &\quad - \frac{1}{2}\sum_{j=1}^{p} \left\{ c_4 + 2b\log\nu_j + 2\frac{b}{\nu_j} \right\},
\end{align}

\noindent where $\{c_i\}_{i=1}^4$ do not depend on $\vect{\nu}$, and

\begin{align} \label{eq:A}
    \vect{A}^{(i)} &= \vect{A}_1^{(i)} - \phi\vect{A}_2^{(i)}\vect{F}' - \phi\vect{F}{\vect{A}^{(i)}_2}' + \phi^2\vect{F}\vect{A}_3^{(i)}\vect{F}' \\
    \label{eq:B}
    \vect{B}^{(i)} &= \sum_{t=1}^T \left[ \left( \vect{y}_t - \vect{X}\vect{\beta}_{t | T}^{(i)} \right) \left(\vect{y}_t - \vect{X}\vect{\beta}_{t | T}^{(i)} \right)' + \vect{X}\vect{V}_{t | T}^{(i)}\vect{X}' \right],
\end{align}

\noindent where

\begin{align}   \label{eq:As}
    \vect{A}_1^{(i)} &= \sum_{t=1}^T \left( \vect{V}_{t | T}^{(i)} + \vect{\beta}_{t | T}^{(i)}{\vect{\beta}^{(i)}_{t | T}}' \right) \nonumber  \\
    \vect{A}_2^{(i)} &= \sum_{t=1}^T \left( \vect{V}_{t,t-1 | T}^{(i)} + \vect{\beta}_{t | T}^{(i)}{\vect{\beta}^{(i)}_{t-1 | T}}' \right) \nonumber \\
    \vect{A}_3^{(i)} &= \sum_{t=1}^T \left( \vect{V}_{t-1 | T}^{(i)} + \vect{\beta}_{t-1 | T}^{(i)}{\vect{\beta}^{(i)}_{t-1 | T}}' \right).
\end{align}

The conditional expectations and covariances in Equation \eqref{eq:As} can be computed with the Kalman Filter~\cite{Kalman:1960kh}, Fixed Interval Smoother~\cite{Rauch:1965wx}, and lag-covariance~\cite{deJong:1988vu} recursions.

\paragraph{The Kalman Filter, Fixed Interval Smoother, and lag-covariance recursions}
\label{par:kffislagcov}

We can use the forward and backward recursions~\cite{Kitagawa:1996p85} to compute the desired expectations and covariances. In the $i$th iteration we initialize the recursion with $\vect{\beta}_{0 | 0}^{(i)} = 0$ and $\vect{V}_{0 | 0}^{(i)} = \vect{C}_0^{(i)}$. For the forward pass, $t=1,\,2,\,\ldots,\,T$, we compute the predictions:

\begin{align} \label{eq:predkf}
    \vect{\beta}_{t | t-1}^{(i)} &= \phi\vect{F} \vect{\beta}_{t-1 | t-1}^{(i)} \nonumber \\
    \vect{V}_{t | t-1}^{(i)} &= \phi^2 \vect{F} \vect{V}_{t-1 | t-1}^{(i)} \vect{F}' + (1-\phi^2) \vect{Q}(\vect{\nu}^{(i)}),
\end{align}

\noindent and filtered estimates:

\begin{align} \label{eq:filtkf}
    \vect{G}_t &= \vect{V}_{t | t-1}^{(i)} \vect{X}' \left(\vect{X} \vect{V}_{t | t-1}^{(i)} \vect{X}' + \vect{I} \right)^{-1}  \nonumber \\
    \vect{\beta}_{t | t}^{(i)} &= \vect{\beta}_{t | t-1}^{(i)} + \vect{G}_t\left( \vect{y}_t - \vect{X}\vect{\beta}_{t | t-1}^{(i)} \right) \nonumber \\
    \vect{V}_{t | t}^{(i)} &= \left( \vect{I} - \vect{G}_t \vect{X} \right) \vect{V}_{t | t-1}^{(i)},
\end{align}

\noindent and for the backward pass, $t=T-1,\ldots,0$, we find the smoothed estimates:

\begin{align} \label{eq:smoothfis}
    \vect{J}_t &= \phi\vect{V}_{t | t}^{(i)}\vect{F}'{\vect{V}_{t+1 | t}^{(i)}}^{-1} \nonumber \\
    \vect{\beta}_{t | T}^{(i)} &= \vect{\beta}_{t | t}^{(i)} + \vect{J}_t\left( \vect{\beta}_{t+1 | T}^{(i)} - \vect{\beta}_{t+1 | t}^{(i)} \right) \nonumber \\
    \vect{V}_{t | T}^{(i)} &= \vect{V}_{t | t}^{(i)} + \vect{J}_t\left(\vect{V}_{t+1 | T}^{(i)} - \vect{V}_{t+1 | t}^{(i)} \right)\vect{J}_t'.
\end{align}

The lag-covariances can be obtained using the covariance smoothing algorithm~\cite{deJong:1988vu}:

\begin{equation} \label{eq:lagcov}
    \vect{V}_{t,t-1 | T}^{(i)} = \vect{V}_{t | T}^{(i)}\vect{J}_{t-1}'.
\end{equation}

The Kalman filter and Fixed Interval Smoother recursions are summarized in Algorithms \ref{alg:kalman} and \ref{alg:fis}, respectively. We should note that the conditional mean and covariance that we ultimately use for source localization (Eq. \ref{eq:empiricalbayesest}) correspond to the Fixed Interval Smoother estimate (FIS, Eq. \ref{eq:smoothfis}) obtained on the final iteration of the MAP-EM algorithm. The FIS provides an estimate of the state based on the full set of measurements, and results in improved performance in terms of reduced error covariance compared to the Kalman Filter (KF), which uses a causal subset of the data~\cite{Anderson:2005tp}.

\begin{algorithm}[t!]
\caption{The Kalman Filter}\label{alg:kalman}
\begin{algorithmic}[0]
\Statex \textbf{Inputs:} $\vect{X}, \{\vect{y}_t\}_{t=1}^T, \phi, \vect{F}, \vect{C}_0, \vect{Q}(\vect{\nu})$

\Statex $\vect{\beta}_{0|0} \gets \vect{0}$ \Comment{Initialization}
\State $\vect{V}_{0 | 0} \gets \vect{C}_0$

\For{$t=1, \ldots T$} \Comment{Kalman's Recursion}
    \State $\vect{\beta}_{t | t-1} \gets \phi\vect{F} \vect{\beta}_{t-1 | t-1}$
    \State $\vect{V}_{t | t-1} \gets \phi^2 \vect{F} \vect{V}_{t-1 | t-1} \vect{F}' + (1-\phi^2) \vect{Q}(\vect{\nu})$
    \State $\vect{G}_t\gets \vect{V}_{t | t-1} \vect{X}' \left(\vect{X} \vect{V}_{t | t-1} \vect{X}' + \vect{I} \right)^{-1}$
    \State $\vect{\beta}_{t | t} \gets \vect{\beta}_{t | t-1} + \vect{G}_t\left( \vect{y}_t - \vect{X}\vect{\beta}_{t | t-1} \right)$
    \State $\vect{V}_{t | t} \gets \left( \vect{I} - \vect{G}_t \vect{X} \right) \vect{V}_{t | t-1}$
\EndFor

\Statex \textbf{Outputs:} $\{\vect{\beta}_{t | t-1}, \vect{\beta}_{t | t}, \vect{V}_{t | t-1}, \vect{V}_{t | t}\}_{t=1}^T$
\end{algorithmic}
\end{algorithm}

\begin{algorithm}[!b]
\caption{The Fixed Interval Smoother}\label{alg:fis}
\begin{algorithmic}[0]
\Statex \textbf{Inputs:} $\{\vect{\beta}_{t | t-1}, \vect{\beta}_{t | t}, \vect{V}_{t | t-1}, \vect{V}_{t | t}\}_{t=1}^T, \phi, \vect{F}$

\For{$t=T-1, \dots, 0$} \Comment{Rauch, Tung, and Striebel's Recursion}
    \State $\vect{J}_t \gets \phi \vect{V}_{t | t}\vect{F}'\vect{V}_{t+1 | t}^{-1}$
    \State $\vect{\beta}_{t | T} \gets \vect{\beta}_{t | t} + \vect{J}_t\left( \vect{\beta}_{t+1 | T} - \vect{\beta}_{t+1 | t} \right)$
    \State $\vect{V}_{t | T} \gets \vect{V}_{t | t} + \vect{J}_t\left(\vect{V}_{t+1 | T} - \vect{V}_{t+1 | t} \right)\vect{J}_t'$
\EndFor

\Statex \textbf{Outputs:} $\{\vect{\beta}_{t | T}, \vect{V}_{t | T}, \vect{J}_t\}_{t=1}^T$
\end{algorithmic}
\end{algorithm}

In Appendix \ref{app:dynvsall}, we show that the Kalman Filter and Fixed Interval Smoother estimates can be interpreted as the solution to a penalized least squares problem, analogous to that of the well-known $L_2$ minimum-norm estimate (MNE)~\cite{Hamalainen:1994uk}. Viewed from this perspective, we can readily see that the FIS, KF, and MNE source localization methods are structurally similar, but with an important difference in how the prior mean for the source activity is represented at a given time. The KF and FIS optimally update their prior means by assimilating data from the past, and, in addition, from the future measurements in the case of FIS. In contrast, the MNE assumes that the prior mean is zero at all times and favors source values close to zero.

\subsubsection{M-step}
\label{subsubsec:mstep}

The M-step in the $i$th iterate is achieved by maximizing with respect to the parameters $\vect{\nu}$ the function $U(\vect{\nu} | \vect{\nu}^{(i)})$ computed in the E-step (Eq. \ref{eq:ufunction}), which equals the sum of two terms: 1) the conditional expectation of the complete data log-likelihood given the full set of measurements and the current estimate of the parameters, and 2) the log-prior density,

\begin{equation} \label{eq:mstep}
    \vect{\nu}^{(i+1)} = \argmax_{\vect{\nu}} U(\vect{\nu} | \vect{\nu}^{(i)}).
\end{equation}

\noindent It is easy to see that the maximum is achieved at

\begin{equation} \label{eq:thetar}
    \nu_j^{(i+1)} = \frac{a^{(i)}_{j,j}\frac{ \lambda \mathrm{tr} (\vect{\widehat{\Sigma}})}{(1-\phi^2)} + 2b}{T + 2b},
\end{equation}

\noindent where $a^{(i)}_{j,j}$ is the $i$th row of the $i$th column of $\vect{A}^{(i)}$ (Eq. \ref{eq:A}), and $b$ is the hyperparameter defined in Equation \eqref{eq:priordensity}.

\subsubsection{Evaluation of the log-posterior density}
\label{subsubsec:evalogpost}

The dMAP-EM algorithm iterates for $i=1,\,2,\,\ldots,\,i_o$ performing E-steps and M-steps until the logarithm of the posterior density of the parameters

\begin{equation} \label{eq:postdens}
    \log \mathrm{Pr}(\vect{\nu} | \{\vect{y}_t\}_{t=1}^T) = \underbrace{\log \mathrm{Pr}(\{\vect{y}\}_{t=1}^T | \vect{\nu})}_{\text{Log-likelihood}}
    + \underbrace{\log \mathrm{Pr}(\vect{\nu})}_{\text{Log-prior}}
    - \underbrace{\log \mathrm{Pr}(\{\vect{y}_t\}_{t=1}^T)}_{\text{Log-evidence}}
\end{equation}

\noindent evaluated at $\vect{\nu} = \vect{\nu}^{(i)}$ reaches an asymptote at some iteration $i_o$.

We can evaluate the logarithm of the likelihood, i.e., the first term in Equation \eqref{eq:postdens}, using the \textit{innovations} form~\cite{Kitagawa:1996p85},

\begin{align} \label{eq:innovations}
    \log \mathrm{Pr}(\{\vect{y}_t\}_{t=1}^T | \vect{\nu}^{(i)}) =&
    - \frac{n T}{2} \log(2\pi) - \frac{1}{2}\sum_{t=1}^T \log\abs{\vect{X}\vect{V}_{t | t-1}^{(i)}\vect{X}' + \vect{I}} \\
    -& \frac{1}{2}\sum_{t=1}^T {\left(\vect{y}_t - \vect{X}\vect{\beta}_{t | t-1}^{(i)}\right)}'\left(\vect{X}\vect{V}_{t | t-1}^{(i)}\vect{X}' + \vect{I}\right)^{-1}
    \left(\vect{y}_t - \vect{X}\vect{\beta}_{t | t-1}^{(i)}\right),
\end{align}

\noindent and the logarithm of the prior is obtained is obtained from Equation \eqref{eq:priordensity}:

\begin{equation} \label{eq:logprior}
    \log \mathrm{Pr}(\vect{\nu}^{(i)}) = \sum_{j=1}^{p} \left\{ c_4 -
    b\log\nu_j^{(i)} - \frac{b}{\nu_j^{(i)}} \right\}.
\end{equation}

\noindent Since the evidence in the data, $\mathrm{Pr}(\{\vect{y}_t\}_{t=1}^T)$, is a constant not depending on $\vect{\nu}$, we do not need to compute it in any iteration.

\subsubsection{Summary of the dMAP-EM algorithm}
\label{subsubsec:dmapemalgor}

The algorithm is initialized with parameters $\vect{\nu}^{(0)}$ and $\vect{C}_0^{(0)}$. In the $i$th iteration, we set the state noise covariance ${\vect{Q}(\vect{\nu}^{(i)}) = [\lambda\mathrm{tr}(\vect{\widehat{\Sigma}})]^{-1}\mathrm{diag}(\vect{\nu}^{(i)})}$ and initial state covariance $\vect{V}_{0 | 0}^{(i)} = \vect{C}^{(i)}_0$. We then perform an E-step (Section \ref{subsubsec:estep}) by running the Kalman Filter, Fixed Interval Smoother, and lag-covariance recursion (Section \ref{par:kffislagcov}), and perform an M-step (Section \ref{subsubsec:mstep}) to update the parameters $\vect{\nu}^{(i+1)}$. At each iteration we can update $\vect{C}_0$ with the heuristic $\vect{C}_0^{(i+1)} = \vect{V}_{0 | T}^{(i)}$. The algorithm iterates for $i=1,\,2,\,\ldots,\,i_o$, performing an E-step followed by an M-step until the log-posterior density evaluated at $\vect{\nu}^{(i_o)}$ converges (Section \ref{subsubsec:evalogpost}). The Maximum a Posteriori (MAP) estimate of the parameters is then $\hat{\vect{\nu}}_{\textsc{map}} = \vect{\nu}^{(i_o)}$, and the Empirical Bayes estimate of the sources amplitudes is $\vect{\beta}_{t | T} = \vect{\beta}_{t | T}^{(i_o)}$. The dMAP-EM algorithm is summarized in Algorithm \ref{alg:dMAPEM}.

\begin{algorithm}[htb]
\caption{The dMAP-EM Algorithm}\label{alg:dMAPEM}
\begin{algorithmic}[0]
\Statex \textbf{Inputs:} $\vect{X}, \{\vect{y}_t\}_{t=1}^T, \phi, \vect{F}, \lambda, \vect{\widehat{\Sigma}}, b$

\Statex \textbf{Initialization:}
\State $i \gets 0$
\State $\vect{\nu}^{(i)} \gets [1, 1, \dots, 1]$
\State $\vect{C}_0 \gets [\lambda\mathrm{tr}(\vect{\widehat{\Sigma}})]^{-1}\vect{I}$

\Repeat
    \State $\vect{Q}(\vect{\nu}^{(i)})\gets [\lambda\mathrm{tr}(\vect{\widehat{\Sigma}})]^{-1}\mathrm{ diag}(\vect{\nu}^{(i)})$

    \Statex \quad \textbf{E-step:}
    \Comment{Algorithms \ref{alg:kalman} and \ref{alg:fis}}
    \State $\{\vect{\beta}_{t | t-1}^{(i)}, \vect{\beta}_{t | t}^{(i)}, \vect{V}_{t | t-1}^{(i)}, \vect{V}_{t | t}^{(i)}\}_{t=1}^T \gets \textrm{KalmanFilter}(\vect{X}, \{\vect{y}_t\}_{t=1}^T, \phi, \vect{F}, \vect{C}_0, \vect{Q}(\vect{\nu}^{(i)}))$
    \State $\{\vect{\beta}_{t | T}^{(i)}, \vect{V}_{t | T}^{(i)}, \vect{J}_t\}_{t=1}^T \gets \textrm{FixedIntervalSmoother}(\{\vect{\beta}_{t | t-1}^{(i)}, \vect{\beta}_{t | t}^{(i)}, \vect{V}_{t |   t-1}^{(i)}, \vect{V}_{t | t}^{(i)}\}_{t=1}^T, \phi, \vect{F})$
    \For{$t=1,\dots, T$} \Comment{Lag Covariance Recursion}
        \State $\vect{V}_{t,t-1 | T}^{(i)} = \vect{V}_{t | T}^{(i)}\vect{J}_{t-1}'$
    \EndFor
    \State $\vect{A}_1^{(i)}\gets \sum_{t=1}^T \left( \vect{V}_{t | T}^{(i)} + \vect{\beta}_{t | T}^{   (i)}{\vect{\beta}^{(i)}_{t | T}}' \right)$
    \State $\vect{A}_2^{(i)}\gets \sum_{t=1}^T \left( \vect{V}_{t,t-1 | T}^{(i)} + \vect{\beta}_{t |    T}^{(i)}{\vect{\beta}^{(i)}_{t-1 | T}}' \right)$
    \State $\vect{A}_3^{(i)}\gets \sum_{t=1}^T \left( \vect{V}_{t-1 | T}^{(i)} + \vect{\beta}_{t-1 |    T}^{(i)}{\vect{\beta}^{(i)}_{t-1 | T}}' \right)$
    \State $\vect{A}^{(i)}\gets \vect{A}_1^{(i)} - \phi\vect{A}_2^{(i)}\vect{F}' - \phi\vect{F}{\vect   {A}^{(i)}_2}' + \phi^2\vect{F}\vect{A}_3^{(i)}\vect{F}'$
    \Comment{Equations \eqref{eq:A} and \eqref{eq:As}}

    \Statex \quad \textbf{Evaluation of $-2 * $ log-posterior:}
    \State $-2 \log \mathrm{Pr}(\{\vect{y}_t\}_{t=1}^T | \vect{\nu}^{(i)}) \gets \sum_{t=1}   ^T \log\abs{\vect{X}\vect{V}_{t | t-1}^{(i)}\vect{X}' + \vect{I}}$ \\
    $\quad \quad \quad + \sum_{t=1}^T {\left(\vect{y}_t - \vect{X}\vect{\beta}_{t | t-1}^{(i)}\right)}   '\left(\vect{X}\vect{V}_{t | t-1}^{(i)}\vect{X}' + \vect{I}\right)^{-1}
        \left(\vect{y}_t - \vect{X}\vect{\beta}_{t | t-1}^{(i)}\right)$
    \State $-2 \log \mathrm{Pr}(\vect{\nu}^{(i)}) \gets 2 b \sum_{j=1}^{p} \left[\log\nu_j^{(i)} + \frac{1}{\nu_j^{(i)}} \right]$

    \Statex \quad \textbf{M-step:}
    \State $i\gets i+1$
    \State $\nu_j^{(i)}\gets \left(a^{(i-1)}_{j,j}\frac{ \lambda \mathrm{tr} (\vect{\widehat{\Sigma}})}   {(1-\phi^2)} + 2b\right)\left(T + 2b\right)^{-1}$

\Until{$-2\log \mathrm{Pr}(\{\vect{y}_t\}_{t=1}^T | \vect{\nu}^{(i)}) - 2 \log \mathrm{Pr}(\vect{\nu}^{   (i)})$ converges}

\State $i_o \gets i-1$
\State $\vect{\hat{\nu}}_{\textsc{map}} \gets \vect{\nu}^{(i_o)}$
\State $\vect{\beta}_{t | T} \gets \vect{\beta}_{t | T}^{(i_o)}$
\State $\vect{V}_{t | T} \gets \vect{V}_{t | T}^{(i_o)}$

\Statex \textbf{Outputs:} $\{\vect{\beta}_{t | T}, \vect{V}_{t | T}\}_{t=1}^T, \vect{\hat{\nu}}_{\textsc{map}}$

\end{algorithmic}
\end{algorithm}

\subsection{Initialization of algorithms}
\label{subsec:inialgor}

The state feedback matrix $\vect{F}$ was constructed as indicated in Section \ref{subsec:spacetimesourcemodel} to incorporate nearest-neighbor interaction in order to model basic local intracortical connections. To evaluate the weights $f_{i,j}$ (Eqs. \ref{eq:invdistance} and \ref{eq:statefeedbackmat}), the distance between dipoles was obtained from the triangular tessellation of the cortical surface. The value of $\phi$ in Equation \eqref{eq:statefeedbackmat} was set to $0.95$: this constrains the modulus of the largest eigenvalue of $\phi\vect{F}$ to be strictly less that $1$, and guarantees stability in the source model (Eq. \ref{eq:statespace}). We should note that although it would be interesting to estimate the parameters $f_{i,j}$ and $\phi$, this would add $\sim 30000$ degrees of freedom to the estimation task---each dipole has about 6 nearest neighbors and there are about 5000 dipoles---making this option unfeasible.

The measurement noise covariance was set equal to the identity matrix since the model had been spatially whitened, i.e., the original raw data model ${\tilde{\vect{y}}_t = \tilde{\vect{X}}\vect{\beta}_t + \tilde{\vect{\varepsilon}}_t}$ was premultiplied by the inverse of a matrix square root of the sample covariance of $\tilde{\vect{\varepsilon}}_t$, which was in turn estimated from from empty room recordings. To initialize the algorithms we set the source covariance at time zero ($\vect{C}_0$) and the input noise covariance as a multiple of the identity:

\begin{equation} \label{eq:initialcov}
    \vect{C}_0 = [\lambda\mathrm{tr}(\vect{\widehat{\Sigma}})] ^ {-1}\vect{I}
\end{equation}

\noindent thus approximating the power signal-to-noise ratio (SNR) of the measurements.

\section{Data analysis}
\label{sec:dataanalysis}

\subsection{Design of simulation studies}
\label{subsec:simulationstudy}

We employed simulation studies to compare the source localization performance of four methods: 1) The $L_2$ minimum-norm estimate (MNE, $\vect{\beta}_t^{\textsc{(mne)}}$ in Equation \eqref{eq:mnevskfvsfisest})~\cite{Hamalainen:1994uk}; 2) An extension of MNE, which we call static Maximum a Posteriori Expectation-Maximization (sMAP-EM), where the variance of each dipole source is estimated similarly to that shown in~\cite{Friston:2008jr} and~\cite{Wipf:2009gj}; 3) The FIS without the estimation of the parameters $\vect{\nu}$ ($\vect{\beta}_{t| T}^{(0)}$ in Equation \eqref{eq:smoothfis}, i.e., the smoothed estimate in the first iteration of our algorithm); and 4) And our dMAP-EM algorithm ($\vect{\beta}_{t| T}^{(i_o)}$ in Equation \eqref{eq:smoothfis}, i.e., the smoothed estimate in the last iteration of our algorithm).

We should note that sMAP-EM uses a static source model and the inverse gamma prior (Eq. \ref{eq:priordensity}) for the source variances. Therefore, the sMAP-EM is similar to our method, but with the state feedback matrix $\vect{F}$ set to zero, i.e., $\vect{F}=\vect{0}$. It provides a baseline to compare how the matrix $\vect{F}$, as specified to model local intracortical connections, can be used as covariates or regressors to explain the source activity $\vect{\beta}_t$ in terms of its past $\vect{\beta}_{t-1}$.

We constructed two simulated data sets with active regions of different sizes to compare the performance of these methods in cases where the activity is distributed across a large area, and where it is highly focal:

\begin{itemize}
    \item \textit{Large Patch:} We selected a large active region within primary somatosensory cortex (Figure \ref{fig:simlarge} top panel);
    \item \textit{Small Patch:} We chose a small active region over primary auditory cortex (Figure \ref{fig:simsmall} top panel).
\end{itemize}

In order to avoid committing an \textit{``inverse crime''}, where simulated data come from the same source space and dynamic system used in estimation, we simulated cortical activation on a highly discretized mesh with $\sim 150,000$ dipole sources in each hemisphere and a temporal generating model differing from that of our autoregressive model. The time course of the sources on each active patch was simulated as a 10-Hz sinusoidal oscillation over a period of one second in order to emulate a realistic MEG experiment, $\sin(2\pi \cdot 10 \cdot \Delta t)$, where the sampling frequency $1/\Delta$ was 200 Hz thus yielding $200$ time samples.

The lead field matrix was computed with the \textit{MNE} software package~\cite{Hamalainen:1989ec} using a single-compartment boundary-element model (BEM) based on high-resolution MRIs processed with \textit{Freesurfer}~\cite{Dale:1999ks,Fischl:1999va} from a human subject. The measurement equation (Eq. \ref{eq:measmodel}) was then used to obtain the simulated MEG recordings, where the measurement noise was set to achieve a power signal-to-noise ratio (SNR) of 5, a value typical for MEG measurements, with signal amplitudes scaled uniformly across the active regions to achieve this SNR.

\subsection{Results of simulation studies}
\label{subsec:simresults}

We compared dipole source estimates and their $95\%$ Bayesian credibility (confidence) intervals (CIs) from the static MNE, sMAP-EM, FIS, and dMAP-EM algorithms, using the simulated measurements described in Section \ref{subsec:simulationstudy}. The $95\%$ CIs are obtained from the diagonal elements of the posterior covariance matrix of each method. We should note that these CIs are Empirical Bayes estimates since we are also conditioning on the model parameters $\vect{\nu}$. Specifically, the $95\%$ CIs of the FIS dipole source estimates are equal to $\vect{\beta}_{t|T}^{(0)}\pm 2$ times the diagonal elements of posterior covariance in the first EM iteration $\vect{V}_{t | T}^{(0)}$ (Eq. \ref{eq:smoothfis}), while the CIs of the dMAP-EM source estimates equal $\vect{\beta}_{t|T}^{(i_o)}\pm 2$ times the diagonal elements of $\vect{V}_{t | T}^{(i_o)}$, i.e. the posterior covariance of the last EM iteration. Similarly, the CIs for the MNE and sMAP-EM estimates can be obtained by using the well-known formula for the posterior covariance in a jointly Gaussian model, or equivalently by setting $\vect{F} = 0$ in our formulation.

\begin{figure}[htb]
    \begin{center}
    \includegraphics{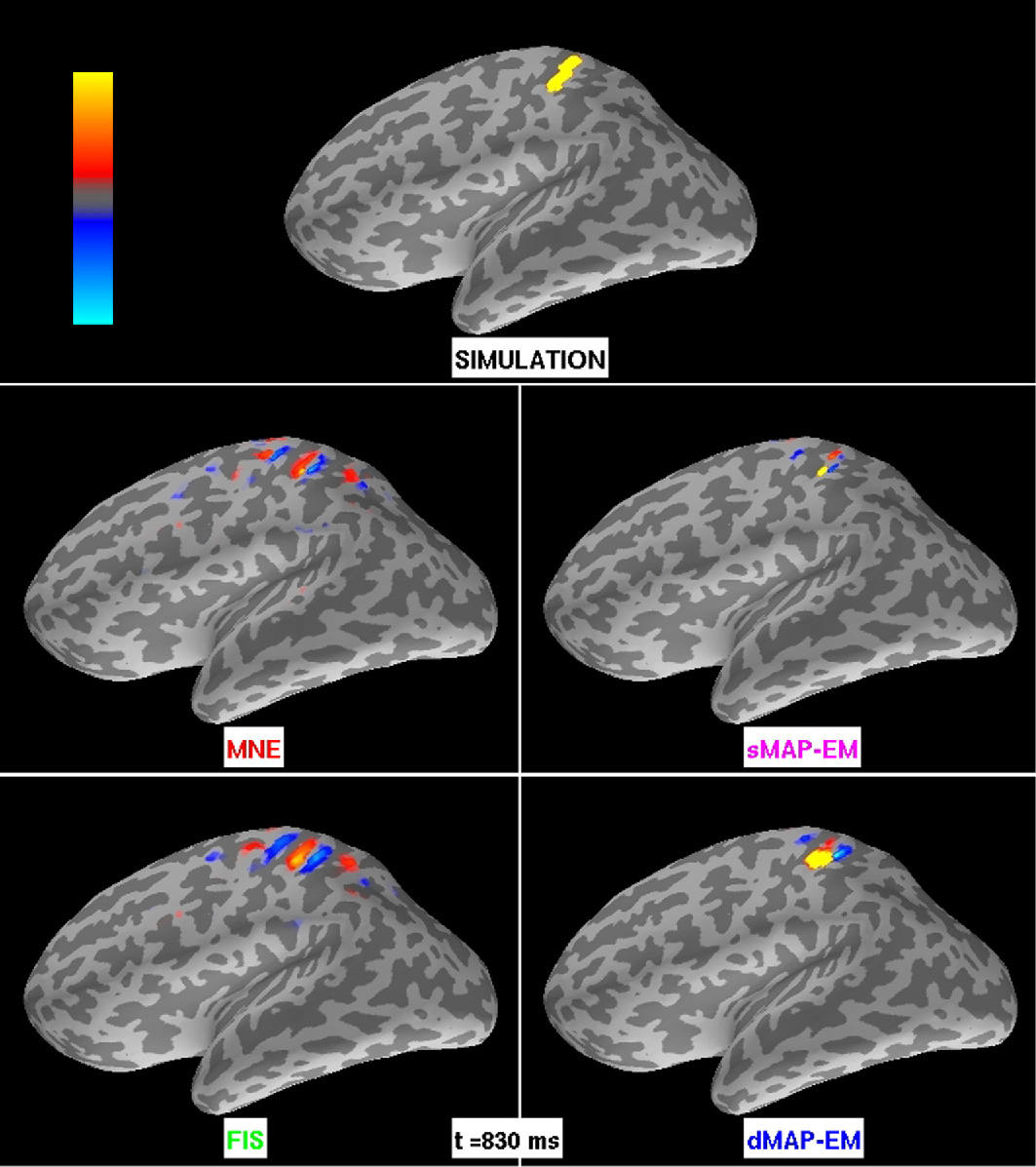}
    \end{center}
    \caption{{\bf Large cortical patch simulation results for MNE, sMAP-EM, FIS, and dMAP-EM methods}. The intensity maps represent the amplitude of simulated and estimated sources. The colorbar's maximum (bright yellow) and minimum (bright blue) corresponds to $\pm$3 nAm for all methods. The top panel shows the simulated activity. The center left panel shows the estimates obtained with MNE, while the bottom left panel shows the FIS estimates. Both of these methods yielded estimates that extended far beyond the simulated active region. The sMAP-EM estimates (center right panel) show a spatial extent significantly smaller than the true activation. The bottom right panel shows the dMAP-EM estimates, whose spatial extent closely matches the simulated active region.}
    \label{fig:simlarge}
\end{figure}

Figure \ref{fig:simlarge} shows the spatial extent of the estimated activity obtained from the large patch simulation. In particular, these intensity maps represent the amplitude of dipole currents $\vect{\beta}_{j,t}$ at a particular time, as opposed to scaled or normalized statistical maps. The MNE extends beyond the simulated area in primary sensory cortex, overlapping pre-central gyrus, parietal cortex, and a small active area in the temporal lobe. The sMAP-EM estimates were highly focal in comparison to the extent of the simulated active region. The FIS estimates were similar to the MNEs, but with better coverage of the simulated patch area. The dMAP-EM method yielded dipole source estimates whose spatial extent closely matched the simulated active region, as observed with the yellow and red hues over primary somatosensory cortex.

\begin{figure}[htb]
    \begin{center}
    \includegraphics[width=\textwidth]{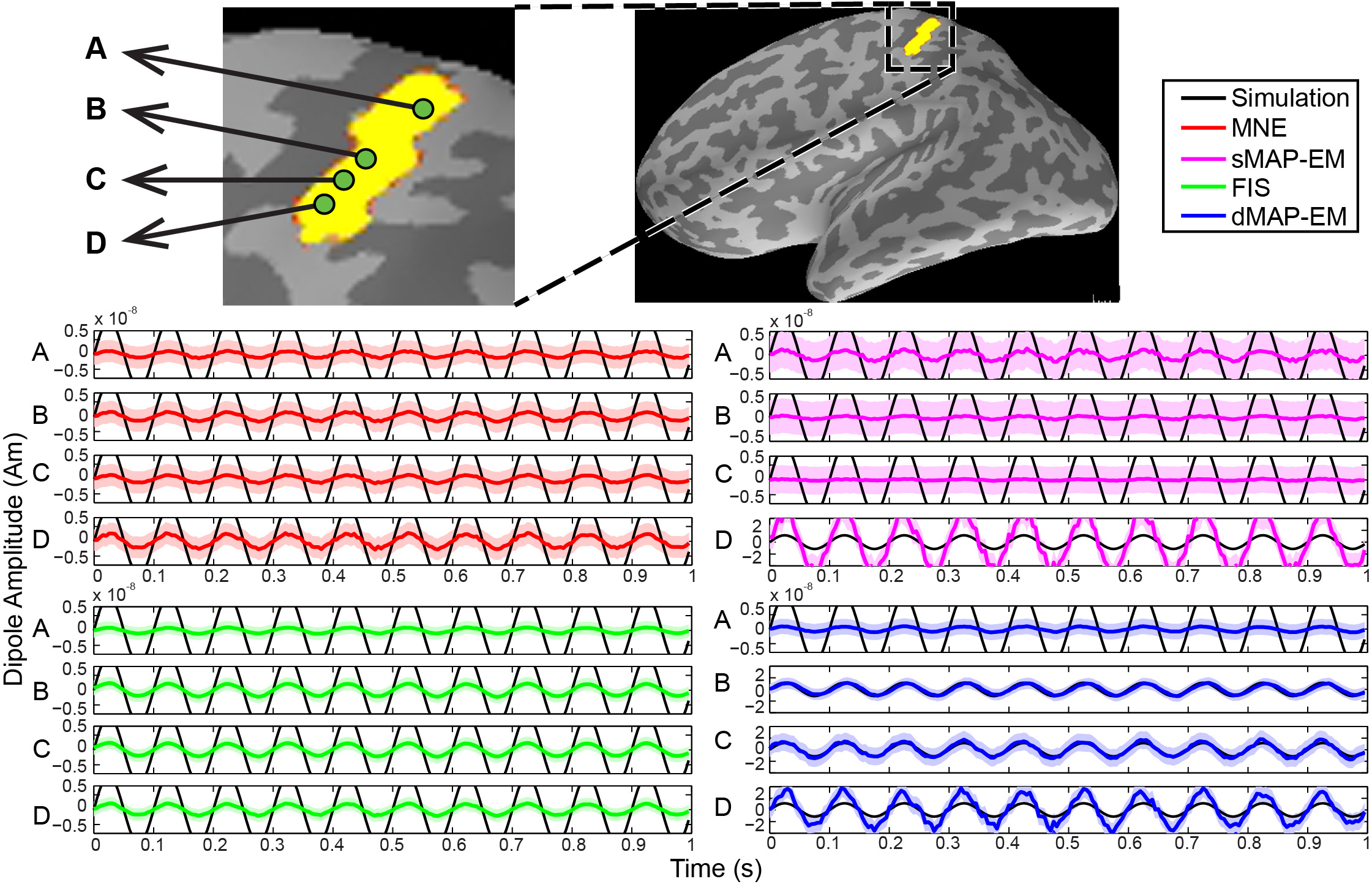}
    \end{center}
    \caption{{\bf Time course estimates for the large patch simulation}. The upper panel shows a zoomed-in view of the simulated cortical area, where green dots represent the selected dipoles labeled A, B, C, and D. The black lines represent simulated sources, while the colored lines represent estimated sources with $95\%$ CIs (colored shading). The center left panel shows the estimated time course of the MNE method in red, and the bottom left panels show the FIS estimates in green. These methods showed poor tracking performance and could not recover the amplitude of the simulated dipole sources. The center right panel shows the sMAP-EM estimates in magenta. This method either underestimated or overestimated the true source amplitude, and showed very large CIs. The bottom right panel shows the estimated time courses for the dMAP-EM method. For 3 of the 4 sources shown (B, C, and D), the dMAP-EM method tracks the simulated time course very closely, significantly better than MNE, FIS, or sMAP-EM, while showing small CIs.}
    \label{fig:simlargetimecourse}
\end{figure}

Figure \ref{fig:simlargetimecourse} compares the temporal tracking performance for representative dipole sources located inside the active area for the large patch simulation. The MNE time series greatly under-estimates the amplitude of the true simulated time series and present wide CIs. The sMAP-EM either underestimates (sub-panels B and C) or overestimates (sub-panel D) the source amplitude and presents larger CIs that the other methods. Similar to MNE, FIS time series estimates also fail to track the true underlying signal accurately, however, they exhibit smaller CIs. For most dipole sources (sub-panels B, C, and D) dMAP-EM closely tracks the true underlying time series while maintaining small CIs. We present in Supplementary Information 1 \href{run:SI1.mov}{(SI1.mov)} a video with the results of this simulation study that visualizes dynamically the spatial localization and temporal tracking performance.

\begin{figure}[htb]
    \begin{center}
    \includegraphics{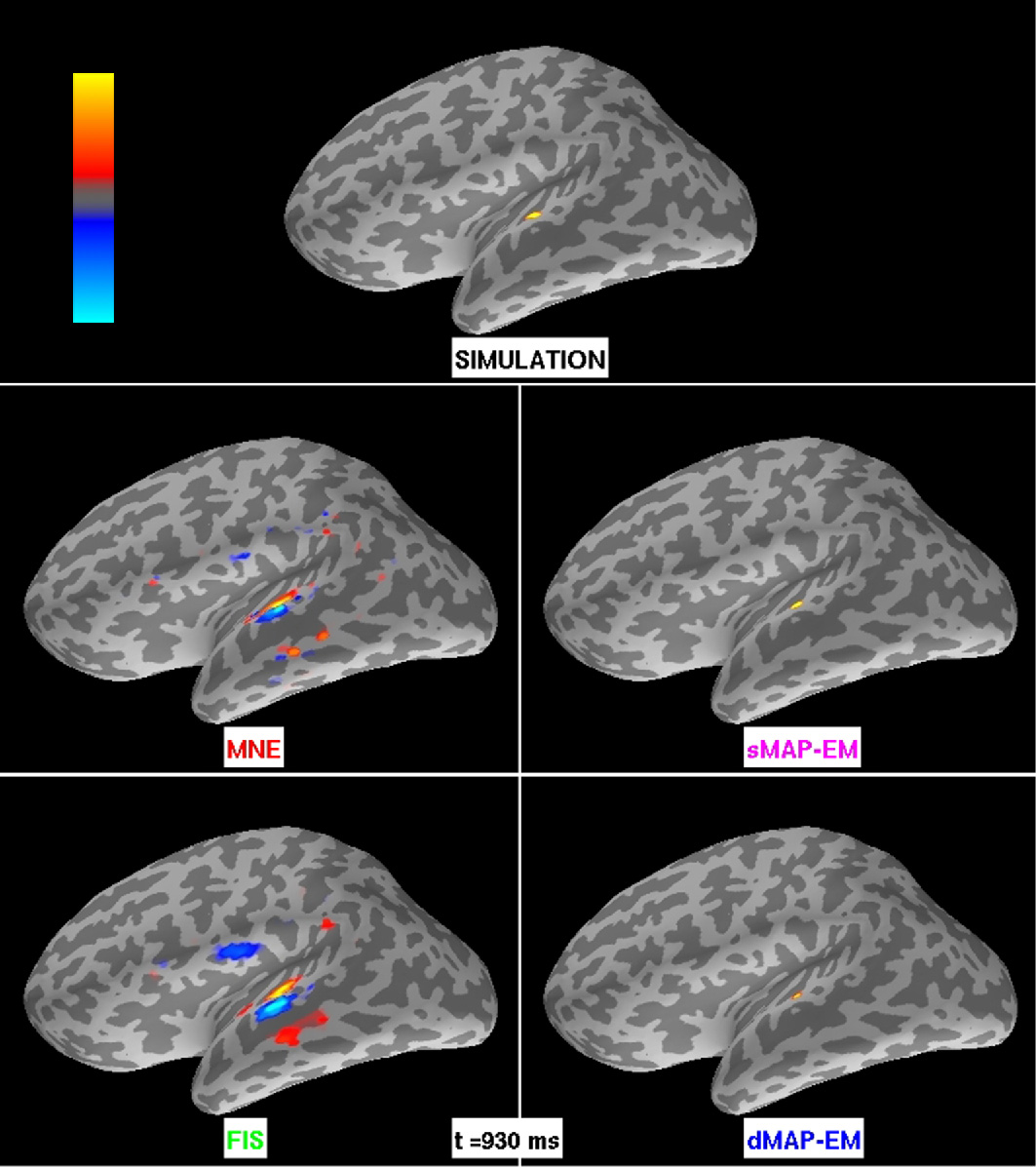}
    \end{center}
    \caption{{\bf Small cortical patch simulation results for MNE, sMAP-EM, FIS, and dMAP-EM methods.} The intensity maps represent the amplitude of simulated and estimated sources. The colorbar's maximum (bright yellow) and minimum (bright blue) corresponds to $\pm$3.5 nAm for MNE and FIS, and $\pm$70 nAm for the simulated data, sMAP-EM, and dMAP-EM method. The top panel shows the simulated activity. The center left and bottom left panels show the estimates obtained with the MNE and FIS methods, respectively. Both methods yield spatially distributed estimates that extend far beyond the true underlying focal active patch. The center right and bottom right panels show the estimates obtained with the sMAP-EM and dMAP-EM methods, respectively. These algorithms yielded focal estimates that closely match the spatial extent of the simulated data.}
    \label{fig:simsmall}
\end{figure}

The estimation results from MNE, sMAP-EM, FIS, and dMAP-EM for the small patch simulation are shown in Figure \ref{fig:simsmall}. Again, the intensity maps represent the amplitude of the source estimates, and not a normalized statistical map. In MNE and FIS the estimates extended beyond the active area to cover regions in the inferior temporal lobe, parietal cortex, and the inferior frontal areas. In contrast, dipole source estimates obtained with sMAP-EM and dMAP-EM are focal and closely match the spatial extent of the active simulated area.

\begin{figure}[htb]
    \begin{center}
    \includegraphics[width=\textwidth]{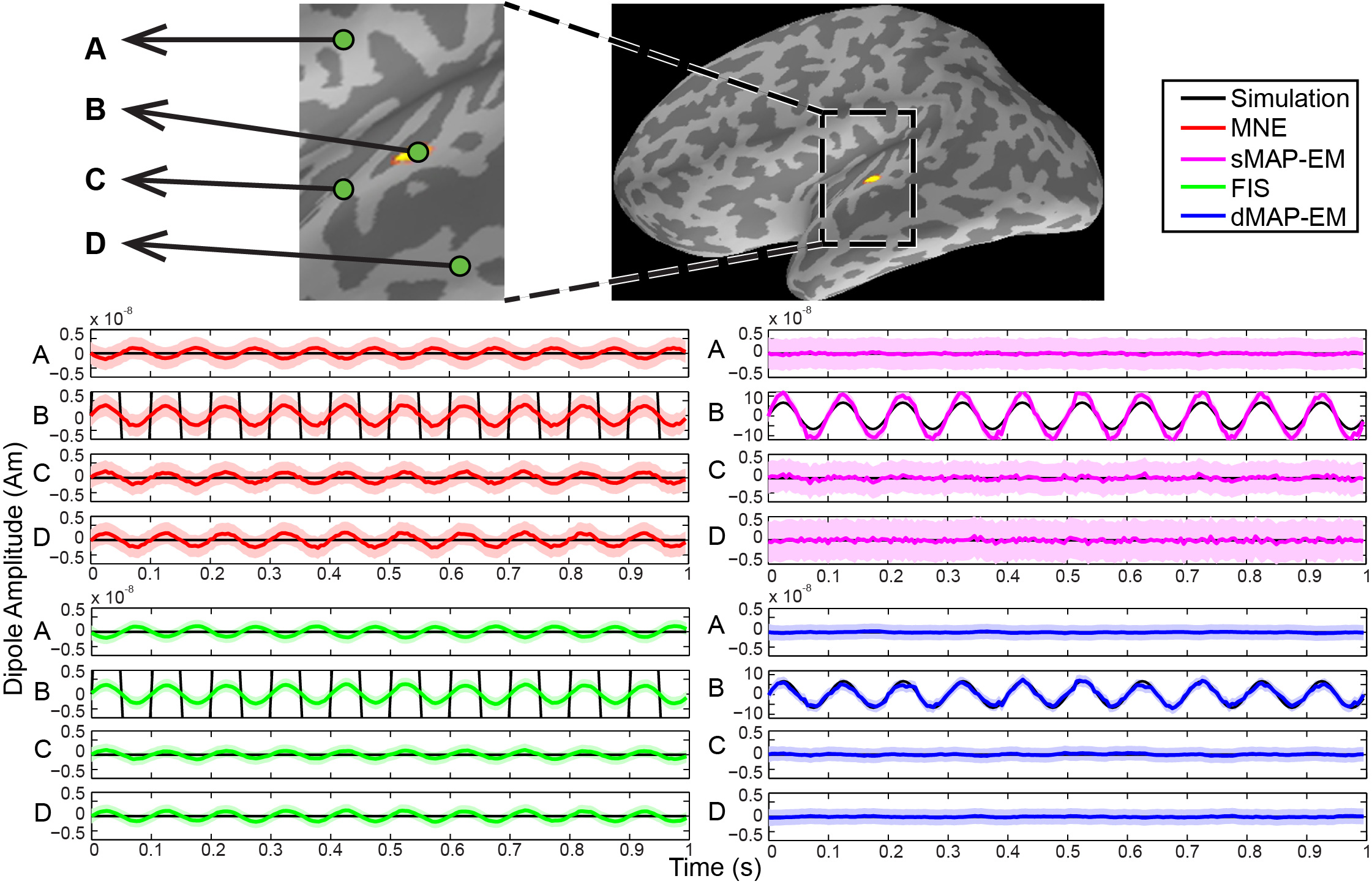}
    \end{center}
    \caption{{\bf Time course of estimation results for the small patch simulation.} The upper shows a zoomed-in view of the simulated cortical area where green dots represent selected dipoles labeled A, B, C, and D. The black lines represent simulated sources, while the colored lines represent estimated sources with $95\%$ CIs (colored shading).The center left panel shows the estimated time courses for the MNE method in red, and the lower left panels show the FIS estimates in green. These methods showed poor tracking performance for the focal active source (sub-panel B), and produced spurious activity in inactive simulated regions (sub-panels A, C, and D). The center right panel shows the sMAP-EM estimates, which tracks the active source (sub-panel B), but presents noisy estimates with very large CIs in the inactive sources (sub-panels A, D, and D). The lower right panel shows the estimated time courses of the dMAP-EM method, which shows accurate tracking performance for the focal active source (sub-panel B), and near-zero activity outside the active area (sub-panels A, C, and D), with very small CIs.}
    \label{fig:simsmalltimecourse}
\end{figure}

Figure \ref{fig:simsmalltimecourse} compares the estimated and simulated time courses for representative dipole sources in the small patch simulation. As shown in the upper left panel, the dipoles labeled A, C and D are outside the active region, while the dipole labeled B is inside the active region. Neither MNE nor FIS are able to accurately track the true underlying active time series (sub-panel B), and both methods produce spurious activation in locations outside the focal active patch (sub-panels A, C, and D). However, the FIS method shows smaller CIs. The sMAP-EM method tracks the active source (sub-panel B), although it slightly overestimates it. In the inactive locations (sub-panels A, C, and D) the sMAP-EM shows small but noisy estimates with very large CIs. The dMAP-EM method accurately tracks the time series for the active dipole source (sub-panel B), while correctly showing small, near-zero activity outside the focal active patch (sub-panels A, C, and D). Furthermore, the dMAP-EM shows small CIs in all cases. We present in the Supplementary Information 2 \href{run:SI2.mov}{(SI2.mov)} a video with the results of this simulation study.

\subsection{Error analyses}
\label{subsec:erroranal}

In this section we evaluate the source localization accuracy of our dMAP-EM methodology in comparison to MNE, sMAP-EM, and FIS. Specifically, we are interested in evaluating the sensitivity and specificity of these source localization methods, as well as the correlation between the simulated sources and their estimates. To do this, we computed receiver operating characteristic (ROC) curves (See~\cite{Moon:2000wn}, or~\cite{Darvas:2004fr} for an application in the MEG/EEG inverse problem) and root mean square errors (RMSE) of sources estimates in the simulation studies described in Sections \ref{subsec:simulationstudy} and \ref{subsec:simresults}.

\subsubsection{Receiver operating characteristic (ROC) curves}
\label{subapp:roccurves}

The ROC technique is used to evaluate the performance of binary detection systems, where the null hypothesis $\mathrm{H}_0$ generally represents the absence of an underlying signal while the alternative hypothesis $\mathrm{H}_A$ denotes its presence. This analysis is done by determining the relation between the detection probability,

\begin{equation} \label{eq:detprob}
    \mathrm{pr}_{D} = \mathrm{Pr}(\text{reject } \mathrm{H}_0 | \mathrm{H}_A \text{ is true}),
\end{equation}

\noindent and the false alarm probability,

\begin{equation} \label{eq:falsealprob}
    \mathrm{pr}_{FA} = \mathrm{Pr}(\text{reject } \mathrm{H}_0 | \mathrm{H}_0 \text{ is true}),
\end{equation}

\noindent as we vary a threshold for the hypothesis test, where in practice, the probabilities are estimated by proportions from simulation studies.

For this analysis we consider a source signal to be inactive ($\mathrm{H}_0$ is true) when the simulated $j$th dipole source at time $t$ equals zero ($\beta_{j,t}^{\textsc{(sim)}} = 0$), while a source signal is considered active ($\mathrm{H}_A$ is true) when $\beta_{j,t}^{\textsc{(sim)}} \neq 0$. For a given estimation method, we define $\hat{\beta}_{j,t}$ as the $j$th dipole source estimate at time $t$. Furthermore, we reject the null hypothesis $\mathrm{H}_0$ to indicate that the source estimate is active with respect to a test threshold $c>0$ if $\abs{\hat{\beta}_{j,t}} > c$. Consequently, for a specific threshold $c$, an estimate of the detection probability $\widehat{\mathrm{pr}}_D$ is given by the fraction of events where we correctly detected an active source, i.e., when the dipole source estimate was considered active ($\abs{\hat{\beta}_{j,t}}>c$) given that the underlying true source was active ($\beta_{j,t}^{\textsc{(sim)}} \neq 0$). Similarly, the estimate of false alarm probability $\widehat{\mathrm{pr}}_{FA}$ is given by the proportion of events where we considered the source estimate to be active ($\abs{\hat{\beta}_{j,t}}>c$) but the underlying true source was inactive ($\beta_{j,t}^{\textsc{(sim)}} = 0$) and made a ``false alarm'' (See Appendix \ref{app:roccurves} for details).

\begin{figure}[htb]
    \begin{center}
    \includegraphics[width=\textwidth]{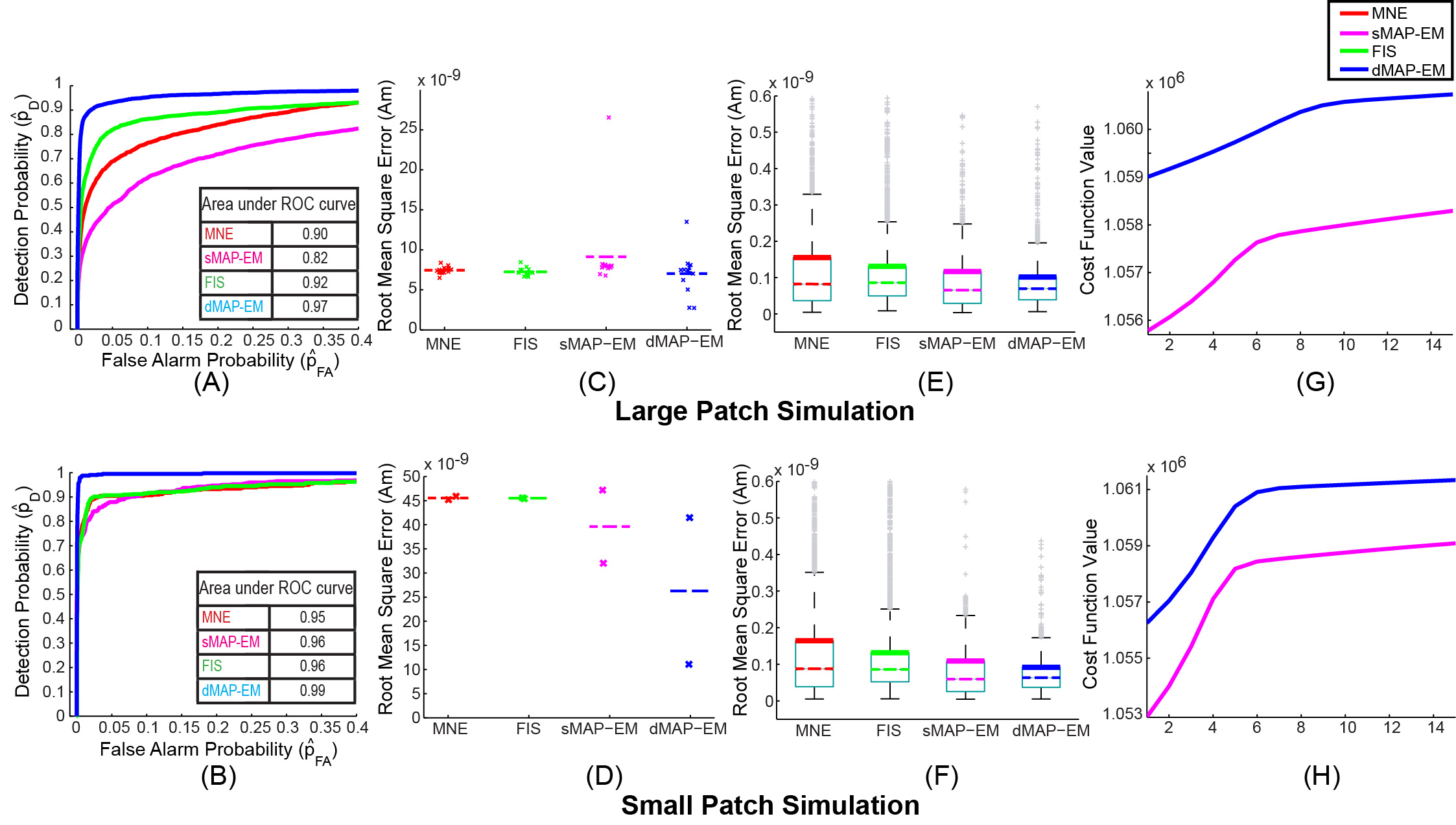}
    \end{center}
    \caption{{\bf ROC curves, RMSE, and convergence of algorithms.} ROC curves and area under the ROC curves from the large patch simulation (sub-panel A) and the small patch simulation (sub-panel B) show that the dMAP-EM method outperforms MNE, FIS, and sMAP-EM methods as it detects more active dipole sources while making significantly less false alarms. Sub-panels C and D show scatter plots of the RMSE of dipoles sources inside the active region in the large patch and small path simulations, respectively. As indicated by the average RMSE (dashed lines), the dMAP-EM method significantly reduces the estimation error compared to the other methods. Sub-panels E and F show box-plot summaries of the RMSE of dipole sources outside the active region from the large patch and small patch simulations, respectively. The dMAP-EM method significantly reduces the $0.99$ quantiles (top horizontal black line or ``whisker'') and $0.75$ quantiles (thick colored line at top of ``box'') of the RMSEs in relation to the other methods. Sub-panels G and H show the convergence of the cost function optimized by the dMAP-EM and sMAP-EM algorithms in the large patch and small patch simulation studies, respectively. In all cases, the cost function reaches a plateau in less than 15 iterations.}
    \label{fig:simerranalconv}
\end{figure}

We computed ROC curves for the large and small patch simulation studies for MNE (red), sMAP-EM (magenta), FIS (green), and dMAP-EM (blue) estimates. The results from the large and small patch simulation study are shown in the sub-panels A and B of Figure \ref{fig:simerranalconv}, respectively. The ROC curves for dMAP-EM showed a superior source detection accuracy in both simulation studies. In the large patch study, the dMAP-EM achieved a $\sim90\%$ detection of true active sources ($\widehat{\mathrm{pr}}_{D} \approx 0.9$) with as few as $\sim2\%$ false alarms ($\widehat{\mathrm{pr}}_{FA} \approx 0.02$), while the other methods required at least $\sim40\%$ false alarms to achieve the same detection accuracy. Similarly, in the small patch simulation, the dMAP-EM achieved a $\sim95\%$ detection of true active sources with as few as $\sim2\%$ false alarms, whereas the other methods required at least $\sim40\%$ false alarms to detect $\sim95\%$ of the true active sources. Furthermore, the area under the ROC curve (see tables in sub-panels A and B of Figure \ref{fig:simerranalconv}), which is a comprehensive measure of the detection accuracy, was greater in the dMAP-EM method than in the other methods, thus indicating a significant improvement in the source localization accuracy of our method.

\subsubsection{Root mean square errors (RMSE)}
\label{subsubsec:rmse}

To further characterize the accuracy of each method, we evaluated the deviation of the dipole source estimates $\hat{\beta}_{j,t}$ in relation to the true simulated sources $\beta_{j,t}^{\textsc{(sim)}}$ in absolute terms by computing root mean square errors (RMSE) in each simulation study. Specifically, for the MNE, FIS, sMAP-EM, and dMAP-EM methods, we computed the RMSE of each dipole source ($j = 1,\,2,\,\ldots,\,p$) over the length in time ($T=200$) of the simulation,

\begin{equation} \label{eq:rmse}
    \text{RMSE}_j = \sqrt{\frac{\sum_{t=1}^T\left(\hat{\beta}_{j,t} - \beta_{j,t}^{\textsc{(sim)}}\right)^2}{T}},
\end{equation}

\noindent and calculated summary statistics of these errors. We separated the summary statistics for the RMSEs of dipole sources inside and outside the simulated active region to disentangle the origin of the estimation errors. Given that the number of dipoles inside the active patch was small, we computed scatter plots and the sample mean of the RMSE of sources in this region (Fig. \ref{fig:simerranalconv}, sub-panels C and D). However, because the number of dipole source located outside the active patch was relatively large, we computed box-plot summaries of the RMSE of sources from this region, thereby displaying $0.01$, $0.25$, $0.5$, $0.75$, and $0.99$ approximate quantiles (Fig. \ref{fig:simerranalconv}, sub-panels E and F). We should note that in box-plots the approximate $0.75$ and $0.25$ quantiles are represented by the bottom (thin line) and top (thick colored line) of the box, while the $0.5$ quantile (median) is denoted by the colored dashed line across the box. The approximate $0.99$ and $0.01$ quantiles are given by the top and bottom ``whiskers'', i.e., the horizontal black lines, and the gray crosses outside whiskers are considered outliers.

Sub-panels C and D of Figure \ref{fig:simerranalconv} show the scatter plots and the average (dash lines) RMSE of dipole sources inside the active region in the large and small patch simulation studies, respectively. In both simulation scenarios the dMAP-EM estimates showed a significant improvement of the average RMSE in relation to the other methods. In the large patch simulation the dMAP-EM method yielded an average RMSE of $7$ nAm, which represents an reduction of $\sim5.4\%$, $\sim2.7\%$, and $\sim23\%$ in the average of errors with respect to MNE, FIS, and sMAP-EM methods, respectively. Similarly, the dMAP-EM method resulted with an average RMSE of $26$ nAm in the small patch simulation showing a reduction of $\sim42\%$ with respect to MNE and FIS, and of $\sim33\%$ in relation to the sMAP-EM errors. In summary, the average RMSE of sources in active regions is significantly reduced in the dMAP-EM method for both large and small patch simulation scenarios.

Sub-panels E and F of Figure \ref{fig:simerranalconv} show the RMSE box-plots of dipole sources located outside the active region in large and small patch simulation, respectively. In both simulated scenarios, our dMAP-EM method yielded the smallest quantiles among the analyzed methods. The reduction in RMSE relative to MNE, FIS, and sMAP-EM are shown in Table \ref{tab:rmseboxplot}. The dMAP-EM method improved the 0.99 and 0.75 RMSE quantiles by 9\% to 51\%, showing that the dMAP-EM estimates significantly and robustly improved source localization accuracy.

\begin{table}[htb]
    \begin{center}
    \begin{tabular}{ c c l |c|c|c|}
        \cline{2-6}
        &   \multicolumn{2}{|c|}{\textbf{\small{Method $\backslash$ Quantile}}} &
        \textbf{0.5} &  \textbf{0.75} &\textbf{0.99}\\
        \cline{1-6}
        \multicolumn{1}{|c|}{\multirow{4}{*}{\begin{sideways}\textbf{Large}\end{sideways}\begin{sideways}\textbf{patch}\end{sideways}}}
                                & \multicolumn{2}{|c|}{dMAP-EM (nAm)}   & 0.06          & 0.10          & 0.19  \\
        \cline{2-6}
        \multicolumn{1}{|c|}{}  & \multicolumn{1}{|l|}{\multirow{2}{*}{Reduction}}
        &       MNE             & 25\%          & 33\%          & 40\%  \\
        \cline{3-6}
        \multicolumn{1}{|c|}{}  & \multicolumn{1}{|c|}{\multirow{2}{*}{with respect to:}} & FIS         & 25\%          & 23\%          & 24\%  \\
        \cline{3-6}
        \multicolumn{1}{|c|}{}  & \multicolumn{1}{|c|}{} & sMAP-EM      & 0\%           & 9\%           & 20\%  \\
        \hline \hline
        \multicolumn{1}{|c|}{\multirow{4}{*}{\begin{sideways}\textbf{Small}\end{sideways}\begin{sideways}\textbf{patch}\end{sideways}}}
                                & \multicolumn{2}{|c|}{dMAP-EM (nAm)}   & 0.06          & 0.09          & 0.17  \\
        \cline{2-6}
        \multicolumn{1}{|c|}{}  & \multicolumn{1}{|l|}{\multirow{2}{*}{Reduction}}
        &   MNE         & 25\%          & 43\%          & 51\%  \\
        \cline{3-6}
        \multicolumn{1}{|c|}{}  &    \multicolumn{1}{|c|}{\multirow{2}{*}{with respect to:}} & FIS          & 25\%          & 30\%          & 32\%  \\
        \cline{3-6}
        \multicolumn{1}{|c|}{}  &   \multicolumn{1}{|c|}{} & sMAP-EM        & 0\%           & 10\%          & 26\%  \\
        \hline
    \end{tabular}
    \end{center}
    \caption[RMSE relative reduction.]{Percentage reduction of the RMSE quantiles for dipole sources outside the active region under dMAP-EM, compared to MNE, FIS, and sMAP-EM}
    \label{tab:rmseboxplot}
\end{table}

\subsection{Convergence and computational requirements}
\label{subsec:convandcomp}

The convergence of the dMAP-EM algorithm was assessed by tracking the logarithm of the posterior density (Eq. \ref{eq:postdens}),

\begin{equation} \label{eq:costfn}
    \text{cost}(\vect{\nu}^{(i)}) = \log \mathrm{Pr}(\{\vect{y}\}_{t=1}^T | \vect{\nu}^{(i)}) + \log \mathrm{Pr}(\vect{\nu}^{(i)})
\end{equation}

\noindent omitting the log-evidence term, which does change during EM iteration. Sub-panels G and H of Figure \ref{fig:simerranalconv} show the convergence, i.e., the cost evaluated at each iteration, of the dMAP-EM and sMAP-EM algorithms for the large patch and small patch simulation studies, respectively. In both simulation scenarios, the cost function reaches a plateau in less than 15 iterations.

The runtime of the dMAP-EM algorithm per EM iteration is effectively $O(Tp^3)$~\cite{Mendel:1971ky}, where we assumed that the number of sensors $n$ is fixed and much smaller than the number of dipole sources $p$ ($n<<p$), and $T$ is the number of measurements in time. The algorithm was implemented in Matlab (The MathWorks, Natick, MA) and run on a dual 6-core Linux workstation at $2.67$ GHz with 24 GB RAM. In our analyses the number of dipole sources was $p = 5124$, the number of sensor was $n=204$, and the number of measurements in time was $T=200$. Since we did not attempt any kind of model reduction procedure, the algorithm yielded a computation time of $\sim 2.5$ hours per EM iteration.

\subsection{Analysis of experimental data from human subjects}
\label{subsec:muresults}

\begin{figure}[htb]
    \begin{center}
    \includegraphics{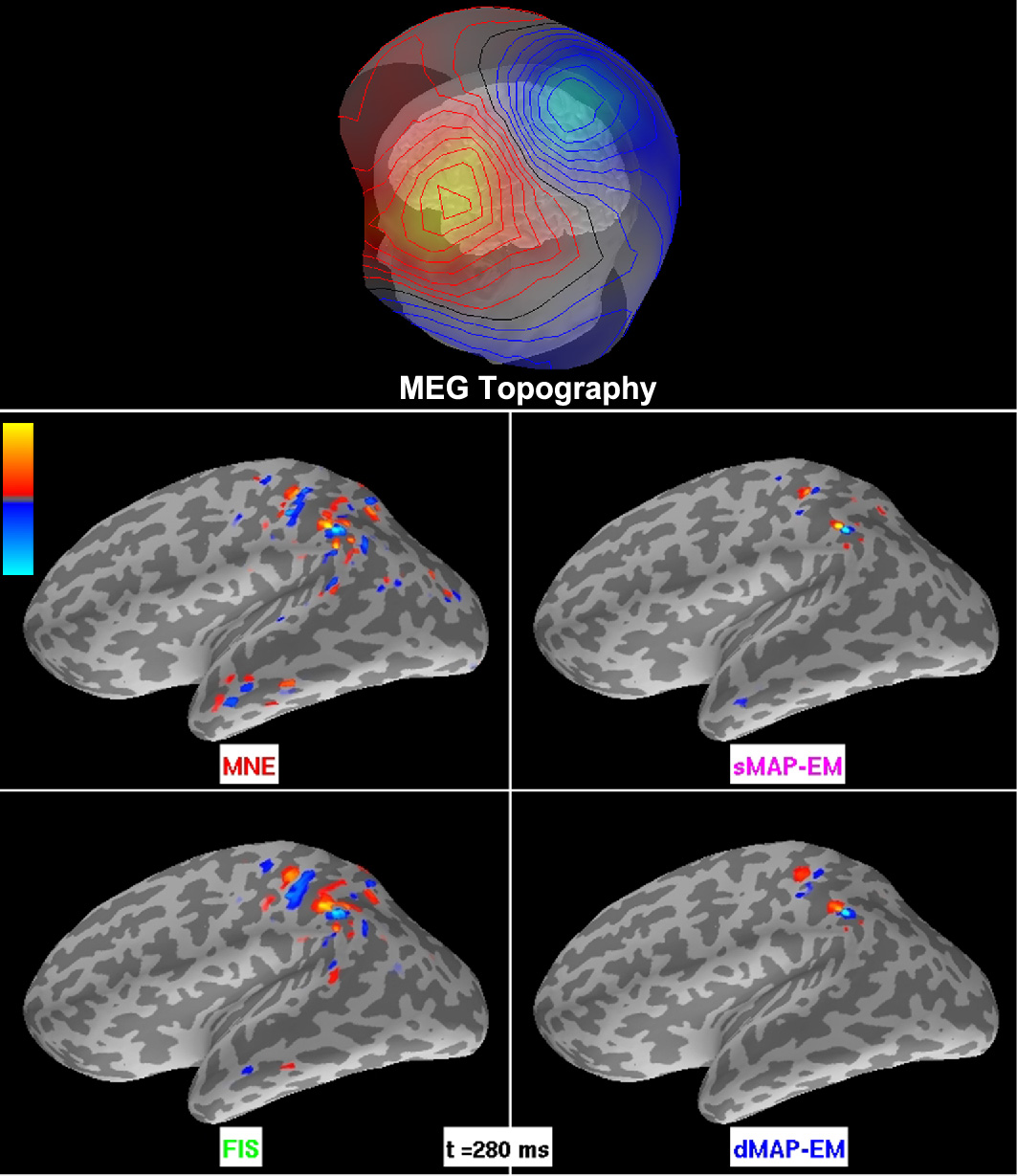}
    \end{center}
    \caption{{\bf Analysis of human MEG \textit{mu}-rhythm data.} The top panel shows the topography of the magnetic field component normal to the sensor surface, with isocontour lines indicating a field change of 100 fT. In the remaining panels, the intensity maps represent the amplitude of source estimates. The colorbar's maximum (bright yellow) and minimum (bright blue) corresponds to $\pm$1.8 nAm for MNE and FIS, and $\pm$9 nAm for the sMAP-EM, and dMAP-EM methods. The center and bottom left panels show the estimates obtained with the MNE and FIS methods, respectively. The estimated activity appears broad and spatially distributed, covering many different cortical regions. Similar to the simulated scenarios, the sMAP-EM algorithm (center right panel) yielded highly focal estimates. The bottom right panel shows the dMAP-EM estimates. This method resulted in more compact estimates covering primary somatosensory cortex and other parietal regions.}
    \label{fig:realmu}
\end{figure}

We also applied the MNE, FIS, sMAP-EM, and dMAP-EM algorithms to \textit{mu}-rhythm MEG data from a human subject. The \textit{mu}-rhythm originates from motor and somatosensory cortices, and consists of oscillations with 10 and 20 Hz components. Data were collected using a 306-channel Neuromag Vectorview MEG system at Massachusetts General Hospital. The subject was instructed to rest with eyes open during the recording. We recorded 12 minutes of data at a sampling frequency of 601 Hz with a bandwidth of 0.1 to 200 Hz, and later downsampled to 200.3 Hz. The data were visually inspected to select 1 second of strong \textit{mu}-rhythm activity, as evidenced by its characteristic ``comb'' shape, for subsequent analyses. In Figure \ref{fig:realmu}, we show the results of the three methods where the intensity maps represent the amplitude of source estimates. This figure also includes the topography of the magnetic field component normal to the sensor surface at the same time instant~\cite{Hamalainen:1994uk,Numminen:1995wd}. Similar to Figures \ref{fig:simlarge} and \ref{fig:simsmall} in the simulated scenario, the MNE and FIS produced broad, spatially distributed estimates with activity covering primary somatosensory and motor regions as well as parietal, occipital, and temporal areas. The sMAP-EM method yielded highly focal estimates that appear spatially irregular. The dMAP-EM estimate presents a more compact spatial extent that covers primary somatosensory and parietal areas.

\begin{figure}[htb]
    \begin{center}
    \includegraphics[width=\textwidth]{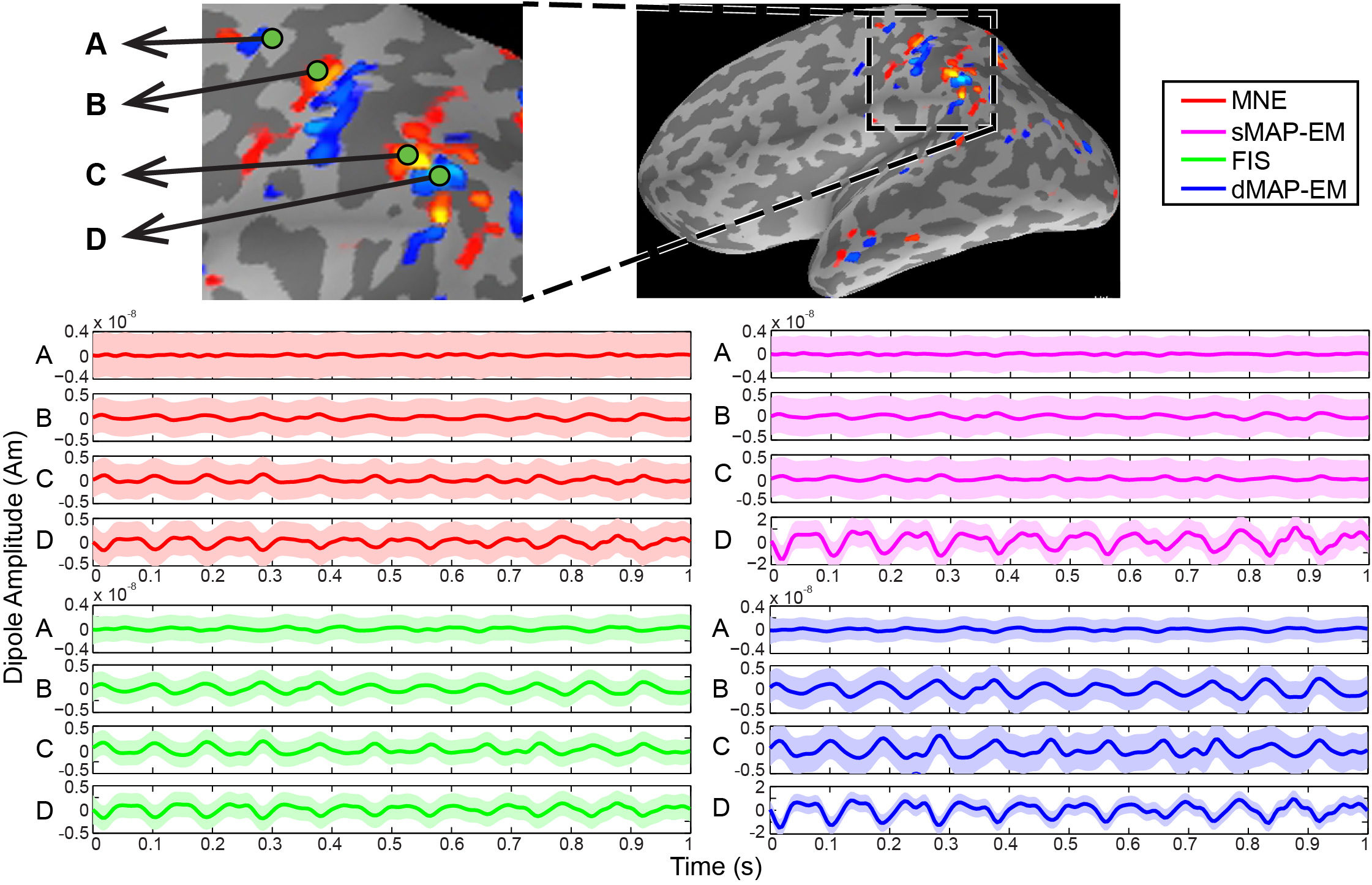}
    \end{center}
    \caption{{\bf Time course of estimation results for human MEG \textit{mu}-rhythm data.} The upper panel shows a zoomed-in view of the cortical activity map obtained with the MNE method, where the green dots represent dipoles labeled A, B, C, and D. The colored lines are estimated sources with $95\%$ CIs represented by light colored areas. The center left panel shows the estimated time course of the MNE method in red, and the bottom left panels show the FIS estimates in green. These methods resulted in estimates with smaller amplitudes. The center right panel shows the sMAP-EM estimates in magenta, which yielded large CIs. The dMAP-EM method (blue) yielded estimates with higher-amplitude oscillations and smaller CIs, with the stereotyped \textit{mu}-shape that characterizes these data (dipoles B, C, and D).}
    \label{fig:realmutimecourse}
\end{figure}

The time course estimates and their $95\%$ Bayesian credibility intervals (CIs) are shown in Figure \ref{fig:realmutimecourse}. The $95\%$ CIs (light colored areas) are obtained as described in Section \ref{subsec:simresults}. The upper panel shows a zoomed-in view of the cortical activity map obtained with the MNE method, where green dots represent the selected dipoles in primary motor (A), primary somatosensory (B), and parietal association areas (C and D). We note that these particular areas were selected as they have been reported to generate the \textit{mu}-rhythm signals~\cite{Hari:1997wa}. Both MNE and FIS methods yielded estimates with smaller amplitudes, however, the CIs for the FIS method were smaller. In sMAP-EM estimates, the amplitude of the dipoles estimates A, B, and C was small with wider CIs in comparison to the dMAP-EM estimates. The dMAP-EM method yielded estimates with higher-amplitude oscillations, especially in dipoles B, C, and D, where the estimate follows the stereotypical ``comb'' shape that characterizes these data, and presented smaller CIs than those of the MNE and sMAP-EM methods. In the Supplementary Information 3 \href{run:SI3.mov}{(SI3.mov)} we present a video of the results of this analysis.

\section{Discussion}
\label{sec:discussion}

Intracranial electrophysiological recordings have shown that on a local scale, brain activity is spatially and temporally correlated~\cite{Bullock:1995wy,Destexhe:1999tc,Leopold:2003va,Nunez:1995vc}. Similarly, non-invasive fMRI and PET studies have shown that brain activation is temporally coherent in a spatially distributed network~\cite{Raichle:2001vf,Gusnard:2001ut,Fox:2005kx,Fox:2007ig}. On the modeling side, biophysical spatiotemporal dynamic models of neuronal networks have been able to simulate electromagnetic scalp signals similar to those seen in recordings during normal and disease states~\cite{Jirsa:2002wz,Wright:2004wl,Robinson:2005dw,David:2005fx,Sotero:2007ij,Kim:2007be,Izhikevich:2008ej,Gross:2001cc}. We incorporated these insights to probabilistically model cortical activation as a distributed spatiotemporal dynamic process, and used this model as the basis for an inverse solution. This probabilistic model acts as a soft \textit{a priori} constraint on the evolution of spatiotemporal cortical trajectories, in a way that allows the recorded data to update our belief on this trajectory at each moment in time. In this model of cortical activity, the nearest-neighbor interactions are intended primarily to model spatial dependencies observed with intracranial electrophysiology. However, as shown by the large and small patch simulation results, the flexibility of this soft constraint results in accurate estimates of both extended and focal brain activity.

Spatiotemporal MEG/EEG source localization algorithms~\cite{Baillet:1997ue,Greensite:2003wn,Daunizeau:2006wl,Daunizeau:2007ji,Friston:2008jr,TrujilloBarreto:2008ck,Zumer:2008in,Limpiti:2009ca,Ou:2009cm,Bolstad:2009eg} have been previously developed with the aim of obtaining temporally smooth estimates by imposing computationally convenient prior constraints on dipole sources. While these methods provide a way of constraining the temporal evolution of inverse solutions, the relationship between prior constraints and underlying physiology is unclear. In contrast, our spatiotemporal dynamic model is structured to represent well-known local cortical biophysics, neuroanatomy, and electrophysiology, and can be developed further to account for more complex brain dynamics and spatial interactions. Other authors have proposed state-space models in the EEG inverse problem with volumetric Laplacian spatial interactions, using recursive least-squares~\cite{Yamashita:2004fb} or an approximate version of the Kalman filter~\cite{Galka:2004ef}. In contrast, our approach establishes dynamic relationships along the \emph{cortical surface} to represent local cortical interactions observed in physiological studies, and uses the full-dimensional Kalman filter and Fixed-Interval Smoother in conjunction with an EM algorithm to provide statistically optimal estimates of dipole source activity as well as dipole-specific model parameters. While these calculations are high dimensional and computationally demanding, we chose this approach in order to place the model in a spatial and temporal scale consistent with the biophysics of cortical interactions.

To understand the dynamic algorithms in relation to previously developed methods, we re-expressed the Kalman Filter and Fixed Interval Smoother estimates in a form analogous to that of the well-known static MNE (see Appendix \ref{app:dynvsall}). This analysis showed that the dynamic methods have a structure that is very similar to MNE, with the Kalman Filter and Fixed Interval Smoother prediction covariance matrices playing the same role as the prior source covariance, or regularization matrix, in MNE. However, there is a critical difference in how the methods account for the prior mean source activity at a given moment in time. MNE assumes that this prior mean is zero at all times, while the Kalman filter and Fixed Interval Smoother optimally update their prior means by assimilating data from the past, as well as the future in the case of FIS. In this way, as pointed out by~\cite{Desai:2005tj} and~\cite{Long:2011fn}, MNE and other similarly structured static algorithms impose a tendency towards zero source activity at every moment in time by ignoring estimates computed for other time points. Any reasonable model approximating spatiotemporal brain dynamics, employed within this dynamic estimation framework, would avoid this tendency towards zero, and would improve performance by enabling information from past and future estimates to help infer the cortical state at a particular point in time.

We designed simulation studies to compare source localization performance of the static MNE, sMAP-EM, FIS, and dMAP-EM estimates. The simulations were constructed to assess algorithm performance on both distributed and focal cortical activity. We simulated MEG data with a highly discretized cortical mesh and a deterministic temporal signal for source activity outside our autoregressive model class to avoid committing an \textit{``inverse crime''}. In both simulations, i.e., with either a large or small active patch, the dMAP-EM method outperformed the MNE, sMAP-EM, and FIS methods in terms of spatial localization accuracy, temporal tracking of the simulated time series, the posterior error covariance, as well as RMSE and ROC analyses. These results suggest that the joint estimation of model parameters and source localization in a spatiotemporal dynamic model, as performed by the dMAP-EM algorithm, can significantly improve inverse solutions. Furthermore, the fact that dMAP-EM algorithm provides more accurate dipole source estimates than FIS, which does not estimate model parameters, indicates that parameter estimation within the dynamic model is critically important in this inverse problem. We also applied these methods to human MEG \textit{mu}-rhythm data, and obtained results similar to the simulated scenarios: The dMAP-EM method yielded distributed yet spatially compact estimates with pronounced time series amplitude in areas of cortex consistent with previous \textit{mu}-rhythm studies, while the MNE and FIS produced spatially spread estimates with small amplitudes, and the sMAP-EM yielded highly focal and spatially irregular estimates.

Recent work by~\cite{Wipf:2009gj} has taken on the important task of characterizing the many seemingly dissimilar static methods described in the MEG inverse literature within a unified statistical framework. An important conclusion of that work is that most static MEG inverse methods can be viewed as solutions to a covariance selection problem, allowing fast covariance selection algorithms from statistics to be applied directly to the MEG inverse problem. Framing the inverse problem in terms of covariance selection, however, leaves open the question of how one should specify the form of that covariance, particularly in the presence of spatiotemporal phenomena. Dynamic modeling and estimation provides a framework to integrate mechanisms and empiricism from biophysics and neurophysiology into solutions for the MEG inverse problem. In this view, the solution to the inverse problem becomes one of specifying and identifying biophysical and dynamical models that closely approximate the underlying neurophysiological system. The spatial and temporal covariance structure then emerges from the second-order statistics inherent in the spatiotemporal dynamic model. If these models can be phrased in the appropriate statistical framework, fast and efficient algorithms with well-known properties can be applied to compute inverse solutions and parameter estimates.

Sophisticated dynamic models have been studied previously in the context of MEG/EEG data~\cite{David:2006gp,Kiebel:2009hq}. These Dynamic Causal Models (DCMs) use simplified spatial representations, such as equivalent current dipoles, to place greater emphasis on temporal modeling while retaining computational tractability. In contrast, the focus of our work has been to explore how spatiotemporal dynamic models in the distributed source framework, and inspired by underlying neurophysiology, can be used to improve source localization. This approach necessitates a higher-dimensional spatial model that can represent local cortical dynamics on a spatial scale consistent with neurophysiological recordings, balanced by a relatively simple temporal model to make computations tractable. In the long run, we envision an approach where more complex spatiotemporal models will be used to solve the MEG/EEG inverse problem. These models could include long-distance connectivity information derived from diffusion tensor imaging (DTI) and nonlinear interactions, while preserving a realistic spatial scale to represent brain activity.

\section{Conclusions and future work}
\label{sec:conclusions}

In this work: 1) we developed a distributed stochastic dynamic model based on a nearest-neighbors autoregression on the cortical surface to represent spatiotemporal cortical dynamics; 2) we derived the dMAP-EM algorithm for optimal dynamic estimation of cortical current sources and model parameters from MEG/EEG data based on the Kalman Filter, Fixed Interval Smoother, and Expectation-Maximization (EM) algorithms; 3) we developed expressions to relate our dynamic estimation method to standard static algorithms; and 4) we applied the spatiotemporal dynamic method to simulated experiments of focal and distributed cortical activation as well as to human experimental data. The results showed that our dMAP-EM method outperforms MNE, sMAP-EM, and FIS methods in terms of spatial localization accuracy, temporal tracking, posterior error covariance, and RMSE and ROC measures.

Our results demonstrate the feasibility of spatiotemporal dynamic estimation in distributed source spaces with several thousand dipoles and hundreds of sensors, resulting in inverse solutions with substantial performance improvements over static methods. Our analysis of known cortical biophysics, models (static vs. dynamic), source estimates, and error in localization revealed clear reasons why one would expect the dynamic methods to perform better than the static MNE. In future work, we will develop new techniques to improve computational performance by means of model reduction or high-performance computing, and will incorporate more realistic neurophysiological models within this dynamic modeling framework.


\setcounter{equation}{0}
\setcounter{figure}{0}
\setcounter{table}{0}
\makeatletter
\renewcommand{\theequation}{A\arabic{equation}}
\renewcommand{\thefigure}{A\arabic{figure}}

\appendices

\section{Robustness of source model against variations in the feedback matrix $\vect{F}$}
\label{app:robustnessmodelF}

In this appendix we analyze how modifications in our spatial model ($\vect{F}$) influence the prior source covariance, and show that the resulting smoothness encoded \textit{a priori} in the dynamic source model is robust against misspecification of the $\vect{F}$ matrix. To simplify our notation, we combine the parameter $\phi$ into the definitions of $\vect{F}$ and $\vect{Q}$. To show the robustness our our model in this area, we first take a modified state model (Eq. \ref{eq:statespace}) where the feedback matrix $\tilde{\vect{F}} = \vect{F} + \vect{\Delta}_F$ has been perturbed by $\vect{\Delta}_F$. Then, we consider the prior source covariance of the original model $\vect{C}$ and that of the modified model $\tilde{\vect{C}}$, when they have reached equilibrium (steady-state), i.e., $\lim_{t\rightarrow\infty} \mathrm{Cov}(\vect{\beta}_t)$. At last, we derive an upper bound for the matrix 2-norm of the difference between the equilibrium covariances $\vect{\Delta}_C = \tilde{\vect{C}}-\vect{C}$, as a function of the perturbation of the spatial model $\vect{\Delta}_F$.

We assume that both $\vect{F}$ and $\tilde{\vect{F}}$ yield stable dynamical systems (i.e., the modulus of their largest eigenvalue is strictly less than 1). From the stability condition, the equilibrium prior state covariance of the original model ($\vect{C}$) and that of the modified model ($\tilde{\vect{C}}$) correspond to the respective (unique) solutions of the discrete Lyapunov equations~\cite{Brockwell:2006tj}:

\begin{align} \label{eq:disclyap}
    \vect{C} = \vect{F}\vect{C}\vect{F}' + \vect{Q} \nonumber \\
    \tilde{\vect{C}}= \tilde{\vect{F}}\tilde{\vect{C}}\tilde{\vect{F}}' + \vect{Q}.
\end{align}

We now express the difference in the equilibrium covariances $\vect{\Delta}_C$ by subtracting the equalities above to obtain:

\begin{equation} \label{eq:deltasigma}
    \vect{\Delta}_C = \vect{F}\vect{\Delta}_C\vect{F}' + \vect{F}\tilde{\vect{C}}\vect{\Delta}_F' + (\vect{F}\tilde{\vect{C}}\vect{\Delta}_F')' + \vect{\Delta}_F\tilde{\vect{C}}\vect{\Delta}_F'.
\end{equation}

\noindent Now we take the matrix induced 2-norm in Equation \eqref{eq:deltasigma}, and apply repeatedly the triangle inequality and the sub-multiplicative property of induced norms:

\begin{equation} \label{eq:normdeltasig0}
    ||\vect{\Delta}_C||_2 \leq ||\vect{F}||_2^2 ||\vect{\Delta}_C||_2 + 2||\vect{F}||_2||\tilde{\vect{C}}||_2||\vect{\Delta}_F||_2 + ||\tilde{\vect{C}}||_2||\vect{\Delta}_F||_2^2.
\end{equation}

\noindent We rearrange terms in Equation \eqref{eq:normdeltasig0} to obtain:

\begin{equation} \label{eq:normdeltasig1}
    ||\vect{\Delta}_C||_2 \leq \frac{||\tilde{\vect{C}}||_2(2||\vect{F}||_2+\vect{\Delta}_F||_2)}{1-||\vect{F}||_2^2}\cdot ||\vect{\Delta}_F||_2
\end{equation}

The fact that both feedback matrices are stable is equivalent to $||\vect{F}||_2,\,||\tilde{\vect{F}}||_2<1$. Using the inverse triangle inequality and setting $\vect{F} = \tilde{\vect{F}} - \vect{\Delta}_F$, we obtain \\${|\,||\tilde{\vect{F}}||_2 - ||\vect{\Delta}_F||_2\,| \leq ||\vect{F}||_2 < 1}$. This inequality implies that $||\vect{\Delta}_F||_2 < 2$. We plug the previous result and the fact that $||\vect{F}||_2<1$ in the inequality above (Eq. \ref{eq:normdeltasig1}), and find that the matrix 2-norm of the difference between the equilibrium covariances $\vect{\Delta}_C$ is bounded above by a constant multiplied by the 2-norm of the perturbation in the feedback matrix $\vect{\Delta}_F$:

\begin{equation} \label{eq:bounddifsteadycov}
    ||\vect{\Delta}_C||_2 < c \cdot ||\vect{\Delta}_F||_2,
\end{equation}

\noindent where the constant $\text{c}=\frac{4||\tilde{\vect{C}}||_2}{1-||\vect{F}||_2^2}$.

From Equation \eqref{eq:bounddifsteadycov} we can conclude that as long as the misspecification of the feedback matrix $\vect{\Delta}_F$ is small, the smoothness modeled \textit{a priori} in the dipole sources is not dramatically altered.

\section{A non-informative prior for the state noise covariance $\vect{Q}(\vect{\nu})$}
\label{app:noninfprior}

In this section we show that setting the parameter $b$ in our prior (Eq. \ref{eq:priordensity}) to a value slightly larger than $3$, makes our prior on $\nu_j$ non-informative (flat). This can be achieved by giving a large variance to the prior while fixing its mode to a value consistent with the order of magnitude in the model. In our case, the order of magnitude of $\nu_j$ ($j\in[1, \dots, p]$) is $1$ since this makes our model consistent with SNR considerations as well as model units. We should note that for $b>3$, the mode of the prior is $1$. This results from the way we parametrize the inverse gamma prior. The variance of our prior is given by:

\begin{equation} \label{eq:varianceprior}
    \mathrm{Var}(\nu_i) = \frac{b^2}{(b-2)^2(b-3)}.
\end{equation}

From this equation (Eq. \ref{eq:varianceprior}) we can see that setting $b=3+\delta$, where $0<\delta<<b$, makes the variance very large and thus give a non-informative (flat) prior.

\section{Derivation of $U(\vect{\nu} | \vect{\nu}^{(i)})$ in the E-step}
\label{app:derestep}

From Equation \eqref{eq:estep} we have that

\begin{equation} \label{eq:estepappendix}
    U(\vect{\nu} | \vect{\nu}^{(i)}) =
    \mathrm{E} \left[\log \mathrm{Pr}\left(
        \{\vect{y}_t\}_{t=1}^T,\{\vect{\beta}_t\}_{t=0}^T | \vect{\nu} \right)
        | \{\vect{y}_t\}_{t=1}^T, \vect{\nu}^{(i)}\right]
     + \log \mathrm{Pr}(\vect{\nu}).
\end{equation}

\noindent where the complete data log-likelihood is derived from the measurement (Eq. \ref{eq:measmodel}) and source (Eq. \ref{eq:statespace}) models:

\begin{equation} \label{eq:compdatalikelihood}
    \log \mathrm{Pr}(\{\vect{y}_t\}_{t=1}^T,\{\vect{\beta}_t\}_{t=0}^T|\vect{\nu})
    = \sum_{t=1}^T \log \mathrm{Pr}(\vect{y}_t | \vect{\beta}_t, \vect{\nu})
    + \sum_{t=1}^T \log \mathrm{Pr}(\vect{\beta}_t | \vect{\beta}_{t-1}, \vect{\nu})
     + \log \mathrm{Pr}(\vect{\beta}_0 | \vect{\nu}),
\end{equation}

\noindent with,

\begin{align} \label{eq:conddensall}
    \log \mathrm{Pr}(\vect{\beta}_0 | \vect{\nu}) &=
    -\frac{1}{2} \{c_{1}+\log|\vect{C}_0| + \vect{\beta}_0'\vect{C}_0^{-1}\vect{\beta}_0\} \nonumber \\
    \log \mathrm{Pr}(\vect{\beta}_t | \vect{\beta}_{t-1}, \vect{\nu}) &=
    -\frac{1}{2} \{ c_{2} + \log|\vect{Q(\vect{\nu})}|
    + (1-\phi^2)^{-1}(\vect{\beta}_t - \phi \vect{F} \vect{\beta}_{t-1})'\vect{Q}(\vect{\nu})^{-1}(\vect{\beta}_t - \phi \vect{F}\vect{\beta}_{t-1}) \} \nonumber \\
    \log \mathrm{Pr}(\vect{y}_t | \vect{\beta}_t, \vect{\nu}) &=
    -\frac{1}{2} \{ c_{3} + (\vect{y}_t- \vect{X}\vect{\beta}_t)'(\vect{y}_t - \vect{X}\vect{\beta}_t) \},
\end{align}

\noindent where $c_1$, $c_2$, and $c_3$ are constants not depending on $\vect{\nu}$.

We apply Equation \eqref{eq:conddensall} to the complete data log-likelihood (Eq. \ref{eq:compdatalikelihood}) and compute its expectation with respect to \\ ${\mathrm{Pr}(\{\vect{\beta}_t\}_{t=0}^T | \{\vect{y}_t\}_{t=1}^T, \vect{\nu}^{(i)})}$, where we have conditioned on the full set of measurements and the parameter estimate of the previous iteration~\cite{Shumway:1982wt}. We then add the logarithm of the prior density of $\vect{\nu}$ and obtain Equation \eqref{eq:ufunction}.

\section{Relationships between the dynamic and static estimators}
\label{app:dynvsall}

In this appendix we present an algebraic analysis of the Kalman Filter (KF) and Fixed Interval Smoother (FIS) estimates that illustrates their relationship to the $L_2$ minimum-norm estimate (MNE)~\cite{Hamalainen:1994uk}. We emphasize that in this analysis the parameters $\vect{\nu}$ in the model are assumed to be fixed. Specifically, we set aside the so-called problem of identification or learning of parameters and focus on comparing and contrasting the functional form of the source amplitude estimates given by these methods.

We note that while it has been well-established that, on one side, the MNE can be seen as a static Maximum a Posteriori estimate of the current dipole sources where the measurements are assumed to be independent in time~\cite{Hamalainen:1994uk,Wipf:2009gj}, and on the other, the KF and FIS algorithms are implementations of a Maximum a Posteriori estimation problem in a linear Gaussian state-space model~\cite{Kitagawa:1996p85,Roweis:1999ul}, our contribution in this section is: 1) To show how the KF, and especially the FIS, are solutions to particular penalized least square problems structurally similar to the $L_2$ minimum-norm cost function, where the penalty term reflects how the information of past $\{\vect{y}_k\}_{k=1}^{t-1}$ and future $\{\vect{y}_k\}_{k=t+1}^{T}$ measurements is optimally accounted by the estimate; and 2) to present the formulas for the KF and FIS estimates in a way that parallel those of the well-known MNE equation as an attempt to further introduce these dynamic estimation techniques in the broad neuroimaging community.

To facilitate the notation, in this section we will assume that all densities are conditioned by a set of parameters and use the notation $\mathrm{Pr}(\cdot) = \mathrm{Pr}(\cdot | \vect{\nu})$. We begin by recalling that the MNE assumes the probability density of the source amplitude vector $\mathrm{Pr}(\vect{\beta}_t)$ to be Gaussian with mean $\vect{\mu}_t = \vect{\mu}^{\textsc{(mne)}} = \vect{0}$ and covariance, or regularization matrix, $\vect{V}_t = \vect{C}^{\textsc{(mne)}}$. Therefore, the MNE maximizes the posterior density~\cite{Hamalainen:1994uk},

\begin{equation} \label{eq:mnepostdens}
    \mathrm{Pr}(\vect{\beta}_t | \vect{y}_t) \propto \underbrace{\mathrm{Pr}(\vect{y}_t | \vect{\beta}_t)}_{\text{Likelihood}}
    \underbrace{\mathrm{Pr}(\vect{\beta}_t)}_{\text{``Prior''}},
\end{equation}

\noindent where the likelihood $\mathrm{Pr}(\vect{y}_t | \vect{\beta}_t)$ is Gaussian with mean $\vect{X}\vect{\beta}_t$ and covariance $\vect{I}$ (Eq. \ref{eq:measmodel}). We must emphasize that the MNE ``prior'' density $\mathrm{Pr}(\vect{\beta}_t)$ does not contain any information about the measurements.

A similar interpretation can be given to the Kalman Filter estimate of $\vect{\beta}_t$ given data up to time $t$, as shown in~\cite{Kitagawa:1996p85,Roweis:1999ul}. In this case, the ``prior'' corresponds to the conditional density of $\vect{\beta}_t$ given data up to time $t-1$, $\mathrm{Pr}(\vect{\beta}_t | \{\vect{y}_k\}_{k=1}^{t-1})$, which is Gaussian with mean $\vect{\mu}_t = \vect{\beta}_{t | t-1}$ and covariance $\vect{V}_t = \vect{V}_{t | t-1}$ given by the prediction step of Kalman Filter recursions (Eq. \ref{eq:predkf}). Then the Kalman Filter computes the Maximum a Posteriori estimate of $\vect{\beta}_t$ given data up to time $t$ by maximizing the posterior density:

\begin{equation} \label{eq:kfpostdens}
    \mathrm{Pr}(\vect{\beta}_t | \{\vect{y}_k\}_{k=1}^t) \propto \underbrace{\mathrm{Pr}(\vect{y}_t | \vect{\beta}_t)}_{\text{Likelihood}}
    \underbrace{\mathrm{Pr}(\vect{\beta}_t | \{\vect{y}_k\}_{k=1}^{t-1})}_{\text{``Prior''}}.
\end{equation}

\noindent We highlight that the Kalman Filter ``prior'' density $\mathrm{Pr}(\vect{\beta}_t | \{\vect{y}_k\}_{k=1}^{t-1})$ contains information only from past measurements $\{\vect{y}_k\}_{k=1}^{t-1}$.

The Fixed Interval Smoother estimate can be interpreted similarly, where the ``prior'' density corresponds to the conditional density of $\vect{\beta}_t$ given previous data up to time $t-1$ and future data after time $t+1$, $\mathrm{Pr}(\vect{\beta}_t | \{\vect{y}_k\}_{k=1}^{t-1}, \{\vect{y}_k\}_{k=t+1}^T)$. To the authors' knowledge, this particular interpretation of the FIS has not been reported in the literature. Again, the ``prior'' density is Gaussian, and its mean $\vect{\mu}_t = \vect{\beta}_{t | T \backslash t}$ and covariance $\vect{V}_t = \vect{V}_{t | T \backslash t}$ can be obtained by standard methods with computationally costly algebraic manipulation. In this case, the notation of the subscript $t | T \backslash t$ reflects that we are conditioning on all data except the immediate data point $\vect{y}_t$. Therefore, the Fixed Interval Smoother estimate maximizes the posterior density of the state given all measurements:

\begin{equation} \label{eq:fispostdens}
    \mathrm{Pr}(\vect{\beta}_t | \{\vect{y}_k\}_{k=1}^T) \propto \underbrace{\mathrm{Pr}(\vect{y}_t | \vect{\beta}_t)}_{\text{Likelihood}}
    \underbrace{\mathrm{Pr}(\vect{\beta}_t | \{\vect{y}_k\}_{k=1}^{t-1}, \{\vect{y}_k\}_{k=t+1}^T)}_{\text{``Prior''}}.
\end{equation}

\noindent We emphasize that the Fixed Interval Smoother ``prior'' density $\mathrm{Pr}(\vect{\beta}_t | \{\vect{y}_k\}_{k=1}^{t-1}, \{\vect{y}_k\}_{k=t+1}^T)$ includes information from both past $\{\vect{y}_k\}_{k=1}^{t-1}$ and future measurements $\{\vect{y}_k\}_{k=t+1}^{T}$.

We can see now that once we have available the ``prior'' densities' mean and covariance for each method, obtaining the Maximum a Posteriori estimates for the MNE, KF, and FIS by maximizing Equations \ref{eq:mnepostdens}, \ref{eq:kfpostdens}, and \ref{eq:fispostdens}, respectively, corresponds to apply a penalized least squares technique where the data fit term corresponds to the likelihood term and the penalty term is related to the ``prior'' density. Specifically, for each method, the penalized least squares function to minimize is

\begin{equation} \label{eq:penleastsq}
        \argmin_{\vect{\beta}_t} \underbrace{||\vect{y}_t-\vect{X}\vect{\beta}_t||^2}_{\text{Data fit}}+
        \underbrace{||\vect{\beta}_t-\vect{\mu}_t||_{\vect{V}_t^{-1}}^2}_{\text{Penalty}},
\end{equation}

\noindent where $\vect{\mu}_t$ and $\vect{V}_t$ are the ``prior'' mean and covariance as defined above for each method. We should note at this point that Equation \eqref{eq:penleastsq} establishes a parallel between $L_2$ minimum-norm cost function, and the Kalman Filter and Fixed Interval Smoother estimates, where the data fit term is identical in these methods, but the penalty varies to indicate how and whether past and future measurements should influence the estimate.

The minimizer of Equation \eqref{eq:penleastsq} is given by

\begin{equation} \label{eq:penleastsqsol}
    \hat{\vect{\beta}}_t(\vect{\mu}_t,\vect{V}_t) = \vect{\mu}_t + \vect{V}_t\vect{X}'(\vect{X}\vect{V}_t\vect{X}'+\vect{I})^{-1}(\vect{y}_t-\vect{X}\vect{\mu}_t).
\end{equation}

\noindent Now we can simply replace the corresponding ``prior'' mean and covariance for each method to obtain the MNE ($\vect{\beta}_t^{\textsc{(mne)}}$), the Kalman Filter estimate ($\vect{\beta}_{t | t}$) and the Fixed Interval Smoother estimate ($\vect{\beta}_{t | T}$):

\begin{equation} \label{eq:mnevskfvsfisest}
    \begin{alignedat}{3}
    \vect{\beta}_t^{\textsc{(mne)}} &= 0    && + \vect{C}^{\textsc{(mne)}}\vect{X}'(\vect{X}\vect{C}^{\textsc{(mne)}}\vect{X}'+\vect{I})^{-1}   &\,\!& (\vect{y}_t-0)           \\
    \vect{\beta}_{t | t}    &= \vect{\beta}_{t | t-1}           &&  +\vect{V}_{t | t-1}\vect{X}'(\vect{X}\vect{V}_{t | t-1}\vect{X}'+\vect{I})^{-1} && (\vect{y}_t-\vect{X}\vect{\beta}_{t | t-1})  \\
    \vect{\beta}_{t | T}    &= \underbrace{\vect{\beta}_{t | T \backslash t}}_\text{Prior mean} &\!\!\!\!&  +\underbrace{\vect{V}_{t | T \backslash t}\vect{X}'%
    (\vect{X}\vect{V}_{t | T \backslash t}\vect{X}'+\vect{I})^{-1}}_\text{Gain}&& \underbrace{(\vect{y}_t-\vect{X}\vect{\beta}_{t | T \backslash t})}_\text{Residual}.
    \end{alignedat}
\end{equation}

Equation \eqref{eq:mnevskfvsfisest} allows us to compare and contrast the algebraic forms of these estimates. We should note that while line 1 is the well-known MNE formula and line 2 is in fact identical to the Kalman Filter algorithm (Eq. \ref{eq:filtkf}), line 3 corresponds to a novel derivation of the Fixed Interval Smoother estimate that allows us to highlight similarities and differences between these estimates. We can see that each method builds a prediction of $\vect{\beta}_t$ using the respective ``prior'' mean, and then updates this prediction using the measurement at the same time point $\vect{y}_t$ by computing a residuals. In the case of MNE, the prediction discards any observed data and assumes it is zero. The Kalman Filter is an improvement over MNE since it builds a prediction based on previous data $\{\vect{y}_k\}_{k=1}^{t-1}$. Finally, the Fixed Interval Smoother goes further to build a prediction based on previous data $\{\vect{y}_k\}_{k=1}^{t-1}$ as well as future data $\{\vect{y}_k\}_{k=t+1}^T$. For all methods, MNE, KF, and FIS, once the prediction is made, the estimate is obtained by updating the prediction using the immediate data point $\vect{y}_t$ by adding a term that is proportional to the residuals.

\section{Computation of ROC curves}
\label{app:roccurves}

In this appendix we present the computations that define the estimates of the detection ($\widehat{\mathrm{pr}}_D$) and false alarm ($\widehat{\mathrm{pr}}_{FA}$) probabilities. These quantities, which depend on the detection threshold $c$, are given by

\begin{align} \label{eq:phat}
    \widehat{\mathrm{pr}}_D(c) =& \frac{\sum_{j=1}^{p}\sum_{t=1}^T \mathrm{indic}(\abs{\hat{\beta}_{j,t}} > c) \cdot
    \mathrm{indic}(\beta_{j,t}^{\textsc{(sim)}} \neq 0)} {\sum_{j=1}^{p}\sum_{t=1}^T \mathrm{indic}(\beta_{j,t}^{\textsc{(sim)}} \neq 0)} \nonumber \\
    \widehat{\mathrm{pr}}_{FA}(c) =& \frac{\sum_{j=1}^{p}\sum_{t=1}^T \mathrm{indic}(\abs{\hat{\beta}_{j,t}} > c) \cdot
    \mathrm{indic}(\beta_{j,t}^{\textsc{(sim)}} = 0)} {\sum_{j=1}^{p}\sum_{t=1}^T \mathrm{indic}(\beta_{j,t}^{\textsc{(sim)}} = 0)},
\end{align}

\noindent where $\mathrm{indic}(\cdot)$ in Equation \eqref{eq:phat} is the indicator function. We should note that $p \approx 5000$ represents the number of dipole sources and $T = 200$ is the number of time samples.

\ifCLASSOPTIONcaptionsoff
  \newpage
\fi

\bibliographystyle{IEEEtran}

\bibliography{main}

\end{document}